\documentclass[aps,pra,amsmath,amssymb,preprintnumbers,twocolumn, showpacs]{revtex4}
\usepackage{amssymb}
\usepackage{color}
\usepackage{graphicx}
\usepackage{subfigure}

\newcommand{\Tr}[1]{\mathrm{Tr} #1}

\newcommand{\DEP}{{\mathrm{DEP}} }
\newcommand{\EB}{{\mathrm{EB}} }
\newcommand{\Phil}{\Phi_{{\lambda}}}
\newcommand{\Phipg}{\Psi_{p,\gamma}}

\begin{document}
\def\bbm[#1]{\mbox{\boldmath$#1$}}

\title{Quantifying the noise of a quantum channel by noise addition}

\author{A. De Pasquale and V. Giovannetti}
\affiliation{NEST, Scuola Normale Superiore and Istituto Nanoscienze-CNR, \\
piazza dei Cavalieri 7, 
I-56126 Pisa, Italy}

\begin{abstract}

In this paper we introduce a way to quantify the noise level associated to a given quantum transformation. The key mechanism lying at the heart of the proposal is \emph{noise addition}: in other words
we compute the amount of extra noise we need to add to the system, 
through convex combination with a reference noisy map or by reiterative applications of the original map,
 before the resulting transformation becomes entanglement-breaking.  We also introduce the notion of entanglement-breaking channels of order $n$
(i.e. maps which become entanglement-breaking after $n$ iterations), and the associated notion of amendable channels (i.e. maps which can be prevented from becoming
entanglement-breaking after iterations by interposing proper quantum transformations). 
 Explicit examples are analyzed  in the context of qubit and one-mode Gaussian channels.

\end{abstract}


\maketitle

\section{Introduction}\label{sec:introduction}

In quantum information several entropic functionals (the so called quantum capacities)  
have been introduced that provide a sort of ``inverse measures"  of the noise level
associated with a given process, see e.g. Refs.~\cite{REV,KEYL,BENSHOR,nielsen}.  
In this approach the  evolution of the system of interest $S$  is described as 
a linear, completely positive, trace preserving mapping (CPT), the quantum channel $\Phi$, which
associates a final state $\Phi(\rho)$ to each possible initial density matrix $\rho$ of $S$. 
The quantum capacities of $\Phi$ have a clear operational meaning as they gauge the optimal communication transmission rates achievable 
when operating in parallel on multiple copies of $S$: consequently the noisier the channel is, the lower are its associated quantum capacities.
Unfortunately however, even for small systems,  these quantities are also extremely difficult to 
evaluate since require optimization over large coding spaces,~e.g. see Ref.~\cite{SHORADD,HASTINGS}.

In this paper we introduce an alternative way to determine how 
disruptive a channel might be which, 
while still having a simple operational interpretation, it is easier to compute than the quantum capacities. 
The starting point of our analysis is to use  Entanglement-Breaking (EB)
channels~\cite{holevoEBT,EBT} as the fundamental benchmarks for evaluating the noise level of a transformation. 
We remind that a map operating on a system $S$, is said to be entanglement-breaking if for all its extensions to an ancillary system  $A$ it annihilates the entanglement between the system and the ancilla \cite{holevoEBT,EBT}. From the point of view of quantum information, the action of these maps represents hence the most disruptive form of noise a quantum mechanical system can undergo. 
A reasonable  way to quantify the noise level of a generic 
map $\Phi$ can then be introduced by computing how much extra noise we need to ``add" to it before the resulting transformation becomes entanglement-breaking. The intuitive idea behind this approach is that  
channels which are less disruptive on $S$ should 
require larger amount of extra noise to behave like an entanglement-breaking map.

 In the following we analyze two different mechanisms of noise addition.
  The first one assumes to form convex combinations 
 $(1-\mu) \Phi + \mu \Phi_\DEP^{\rho_0}$
 of the input channel $\Phi$ with 
 generalized completely depolarizing channels $\Phi_\DEP^{\rho_0}$ (these are the most drastic examples of entanglement-breaking channels
which bring every state of $S$ into a unique output configuration $\rho_0$ -- the fixed point of the map).
In this approach the level of noise associated with  the original map $\Phi$ is gauged by the minimum value 
$\mu_c$
of the mixing parameter $\mu$ which transforms $(1-\mu) \Phi + \mu \Phi_\DEP^{\rho_0}$ 
 into an entanglement-breaking  map (as we will see a proper characterization of this measure requires an optimization upon~$\rho_0$). 
 
 The second mechanisms of noise addition we consider  assumes instead the reiterative application of $\Phi$ on $S$.  In this case the noise level is determined by the minimum value  $n_c$ of
 iterations   needed to transform $\Phi$ in an entanglement-breaking map ({\em if} such minimum exists). As we shall see,   due to the lack of monotonicity under concatenation  with other maps,
  this second functional cannot be considered a proper measure of the noise 
 level introduced by 
 $\Phi$ 
  (regularized version of $n_c$ \emph{do} however retain this property).
 Nonetheless  $n_c$ 
 captures some important aspects of the dynamics 
 associated with $\Phi$: namely it counts 
 the number of discrete time evolutions induced by the map that a system can sustain before  its entanglement  with  an external ancilla is completely  destroyed.  The definition of $n_c$ 
      gives us also  the opportunity of introducing the set  $\EB^n$ of the entanglement-breaking channels of order $n$, and the notion of \emph{amendable} channels.
    The former  is composed by  all  CPT maps $\Phi$ which, when applied $n$ times, are entanglement-breaking.  Vice-versa a channel $\Phi$ is amendable if it can
    be prevented from becoming entanglement-breaking after $n_c$ iterations via a proper application of intermediate quantum channels -- a similar problem was also discussed in Ref.~\cite{gavenda}.

  In the following we will discuss some general
 features of  the functional  $\mu_c$ and $n_c$ computing their exact values for some special class of channels. 
Specifically 
 in Sec. \ref{sec:critmu} we introduce the functionals and characterize some general properties. 
 In Sec.  \ref{sec:qubit} we will restrict our attention to qubit channels computing the value of $\mu_c$ and $n_c$  for  the set of  unital maps and for  the generalized amplitude-damping channels. 
 Also using examples from these sets we will show that the set $\EB^n$ is not convex for $n\geq 2$ and provide evidence of  the existence of amendable maps. In Sec.  \ref{sec:gaussian} we will consider the analogous of qubit maps for continuous variable systems, that is one-mode gaussian channels, and in particular we will evaluate the functional $n_c$ for attenuation and amplification channels. Conclusions and final remarks are given in Sec.~\ref{par:Summary and Conclusions}.

 \section{Definitions and basic properties}\label{sec:critmu}

It is a well known \cite{holevoEBT,EBT} fact that any  entanglement-breaking channel $\Phi_{\EB}$  can be described  as measure and re-prepare scheme according to which 
 the system $S$ is  initialized  in some state $\rho_\ell$ depending on  the  outcome $\ell$ of a measurement performed on the input state $\rho$, i.e.
\begin{equation}
\Phi_{\EB}[\rho]=\sum_{\ell} \rho_\ell \Tr[F_\ell\rho]\;,
\end{equation}
(here  $\{F_\ell\}$ are the elements of the   positive operator valued measure (POVM) which defines the measuring process).
These maps cannot be used to convey quantum information even in the presence of two-way classical side channel (of course  they might allow to achieve this goal when the communicating parties are given some prior shared entanglement via teleportation). In this respect they can thus be considered
as ``classical" communication lines -- notice however that they might be used to 
create {\em quantum  discord}~\cite{DISC}.  
One can easily verify that the set EB formed by the entanglement-breaking channels is closed under  convex combinations and under sequential concatenation, i.e. 
given $p\in[0,1]$ and $\Phi_{\EB}'$, $\Phi_{\EB}'' \in \EB$ we have $p \Phi_{\EB}' + (1-p) \Phi_{\EB}''\in \EB$ and $\Phi_{\EB}'\circ \Phi_{\EB}''\in \EB$ [here
``$\circ$" stands for  the composition of super-operators defined in Sec.~\ref{moncon}]. More generally,
it also true that the set $\EB$ is stable under concatenation with other (non necessarily entanglement-breaking) maps, i.e. 
\begin{eqnarray}
\Phi_{\EB} \in \EB \Longrightarrow \Phi\circ \Phi_{\EB},  \Phi_{\EB}\circ \Phi \in \EB \;, \quad  \mbox{$\forall\; \Phi$ CPT.}
\label{impoNEW} 
\end{eqnarray}

\subsection{Noise addition via convex convolution} \label{sec:convex}

A proper subset of  the EB channels is provided by the completely depolarizing maps, which transform any input state $\rho$ of $S$ into an assigned
fixed point  $\rho_0\in \mathfrak{S}({\cal H}_S)$,~i.e. 
\begin{equation}
\Phi^{\rho_0}_\DEP [\rho]= \rho_0 \Tr[\rho]\;. 
\end{equation}
(we indicate with $\mathfrak{S}({\cal H}_S)$ the set of density matrices of~$S$).
In a sense, these maps form the hard core of the EB set which prevents any sort of communication (not even classical). We can thus use them as 
a fundamental ``unity" of added noise. 
Let then $\Phi$ be a generic map acting on $S$. For each completely depolarizing channel
$\Phi^{\rho_0}_\DEP$ we define $\mu(\Phi; {{\rho_0}})$ 
to be the minimum value of the mixing probability parameter $\mu\in [0,1]$ that transforms the convex convolution 
$(1-\mu) \Phi + \mu \Phi^{\rho_0}_\DEP$ into an element of EB,~i.e.  
 \begin{equation}\label{eq:semimuc}
 \mu(\Phi; {{\rho_0}}):=\min_{\mu \in [0,1]} \left\{(1-\mu) \Phi + \mu \Phi^{\rho_0}_\DEP \in \mbox{EB}\right\}\;.
 \end{equation}
 Thanks to the  Choi-Jamiolkowski isomorphism~\cite{JAM}, computing $\mu(\Phi; {{\rho_0}})$ corresponds to  determining the minimum $\mu$ for which the state 
 \begin{equation}\label{eq:CJ}
\Gamma_{\rho_0,\mu}^{\Phi}=(1-\mu) (\Phi \otimes I) [\psi_{+}]+\mu \; \rho_0  \otimes  \frac{\openone}{d}
\end{equation}
is separable. Here we used the symbol  $\psi_{+}$ to represent the  density matrix  $|\psi_{+}\rangle\langle \psi_{+}|$ associated with 
the maximally entangled state
$|\psi_{+}\rangle= \frac{1}{\sqrt{d}} \sum_{j=1}^{d} |j\rangle  \otimes  |j \rangle \in {\cal H}_S^{\otimes 2}$, 
$\{|j\rangle\}$ being an orthonormal basis of  the Hilbert space of the system $S$, and
$d=\dim{\mathcal{H}_{\mathcal{S}}}$ being its dimension. We remind  that the specific choice of such basis is completely irrelevant when constructing the
Choi-Jamiolkowski state.

 The quantity~(\ref{eq:semimuc}) measures the amount of $\Phi^{\rho_0}_\DEP$ we need to mix  to $\Phi$ via a classical 
 stochastic process   in order to make the resulting map entanglement-breaking. Clearly 
 $\mu(\Phi; {{\rho_0}})$ nullifies if $\Phi$ is already an element of EB.
Vice-versa for non entanglement-breaking channels $\Phi$ the value of $\mu(\Phi; {{\rho_0}})$ is always non null and in general might depend on the selected $\Phi^{\rho_0}_\DEP$ 
(some of the properties of $\mu(\Phi; {{\rho_0}})$ are discussed in Appendix~\ref{appendixA}).
To get a functional  of $\Phi$ alone we need hence to optimize with respect to the possible choices of the fix point 
$\rho_0$. This brings us to define the function
\begin{equation}\label{eq:muc}
\mu_c(\Phi)  :=\min_{{\rho_0}} \mu(\Phi; {{\rho_0}})\;,
\end{equation}
which constitutes our first (inverse) measure of the noise level associated with $\Phi$.  Solving the minimization
in Eq.~(\ref{eq:muc})  is not simple in general:  still in the next section we shall see that for some classes of channels which possess special symmetries this is feasible. 
Here we present some  basic properties of the functional $\mu_c(\Phi)$.  

\subsubsection{Monotonicity under Concatenation}\label{moncon}
Given $\Phi$ and $\Psi$ CPT transformations we have 
\begin{eqnarray}
\mu_c (\Phi \circ \Psi)&\leq&  \mu_c (\Psi)\;, \label{prop3}  \\
\mu_c (\Phi \circ \Psi)&\leq&  \mu_c (\Phi)\;, \label{prop4}
 \end{eqnarray}
 with  ``$\circ$" being the composition of super-operators such that for all $\rho$  we have
 $(\Phi\circ\Psi)[\rho]=\Phi[\Psi[\rho]]$.
Property~(\ref{prop3}) is almost straightforward. Indeed for all $\mu \geq \mu_c(\Psi)$ there exists a density matrix $\rho_\Psi$ such that $ \Gamma^{\Psi}_{\rho_{\Psi},\mu} $ is separable.
Since for every $\Phi \in$ CPT the state $(\Phi \otimes I)  \Gamma^{\Psi}_{\rho_{\Psi},\mu}$ is still separable, $\mu\geq\mu_c(\Phi\circ \Psi)$ which implies $\mu_c(\Phi \circ \Psi)\leq \mu_c(\Psi)$.
Property~(\ref{prop4})  can be proved analogously. Indeed if $\mu\geq\mu_c(\Phi)$ there exists a depolarizing channel $\Phi_{\DEP}^{\rho_\Phi}$ such that
\begin{equation}
(1-\mu)(\Phi \otimes I) + \mu (\Phi^{\rho_{\Phi}}_{\DEP} \otimes I) \in \EB.
\end{equation}
Thus, if we apply this map to the state $(\Psi \otimes I)(\psi_+)$ we get a separable state, and property~(\ref{prop4}) is immediately proved.

An almost trivial  consequence of the above inequalities is  the fact that $\mu_c$ is constant under unitary transformations, i.e. 
\begin{eqnarray}\label{eq:mucinvariance}
\mu_c ({\cal U} \circ \Phi \circ {\cal V} )&=&  \mu_c (\Phi)\;, \label{prop5}  
 \end{eqnarray}
where ${\cal U}$ and  ${\cal V}$ stands for unitary channels, i.e. ${\cal U}(\rho):= U\rho U^\dag$ with $U$ being unitary. 
\subsubsection{Extremal values}\label{extremal}
By definition $\mu_c(\Phi)$ is always positive (it nullifies if and only if $\Phi\in \EB$) and smaller than~$1$. 
As a matter of fact, it is possible to show that
\begin{equation}
\mu_c(\Phi) \leq  \label{UPPERBOUND}
 \frac{d}{1+d}\;,
\end{equation}
the upper bound being achieved for noiseless (unitary) channels. To see this we first 
use the fact that 
the generalized Werner states
\begin{equation}
\Gamma_{\openone/d,\mu}^{I} := (1-\mu) (I\otimes I) [\psi_{+}] + \mu\;  \frac{\openone}{d} \otimes \frac{\openone}{d}  
\end{equation}
are separable  iff $\mu \geq d/(1+d)$~\cite{WERNER}.  Therefore since $\Gamma_{\openone/d,\mu}^{I}$ is the
Choi-Jamiolkowski state of the identity channel $\Phi=I$ this implies,
  \begin{eqnarray}
 \mu(I; {{\openone/d}} ) =  d/(1+d) \label{identity1}\;.
 \end{eqnarray}
 Now the bound~(\ref{UPPERBOUND}) can be easily established by noticing that 
\begin{eqnarray}
\mu_c(\Phi)&\leq&  \mu(\Phi;  {{\Phi(\openone)/d}})
\leq   \mu(I; {{\openone/d}})
=  d/(1+d)\;, \nonumber \\
\end{eqnarray} 
where the first inequality is just a consequence of the definition~(\ref{eq:muc}) while the  second one 
follows from the fact that the Choi-Jamiolkowski state  
$\Gamma_{\Phi(\openone)/d,\mu}^{\Phi}$ can be obtained from 
$\Gamma_{\openone/d,\mu}^{I}$ via the application of a local channel, i.e. 
$ \Gamma_{\Phi(\openone)/d,\mu}^{\Phi} =( \Phi\otimes I )[\Gamma_{\openone/d,\mu}^{I}]$.
Remains to show that the threshold value~(\ref{eq:muc})  is achievable for unitary channels, i.e. 
\begin{eqnarray}
\mu_c({\cal U})=d/(1+d) \label{boundachi}\;.
\end{eqnarray}
Thanks to the property~(\ref{prop5}) we can just focus on the case ${\cal U}=I$ for which we have already established the 
condition~(\ref{identity1}).  To prove the relation~(\ref{boundachi})
 we need hence only to verify that $\mu(I;{{\openone/d}}) \leq \mu(I;{{\rho_0}})$ for all $\rho_0$. 
This can be proven by observing that $\Gamma_{\openone/d,\mu}^{I}$ is obtained from $\Gamma_{\rho_0,\mu}^{I}$
 via the application of 
 the twirling isotropic map~\cite{dist,TWIRL} 
\begin{eqnarray}
{\cal P}[\cdots] := \int  dm(V) ({ V} \otimes { V}^*) [\cdots] ({ V} \otimes { V}^*)^\dag\;, 
\end{eqnarray} 
where the integral is evaluated over the elements $V$ of the unitary group on ${\cal H}_S$, $dm(V)$ is the Haar measure, while $V^*$ represents
the unitary transformation obtained from $V$ by complex conjugation with respect to the canonical basis $\{ |j\rangle\}$.
 Indeed, on one hand $|\psi_+\rangle$ is invariant under 
the application of $V\otimes V^*$ (i.e. 
 $V\otimes V^* |\psi_+\rangle = |\psi_+\rangle$) while on the other hand $\int dm(V) V \rho_0V^\dag = \openone/d$, so that
\begin{eqnarray}
{\cal P}(\Gamma_{\rho_0,\mu}^{I}) = \Gamma_{\openone/d,\mu}^{I}\;.
\end{eqnarray}
 Take then $\mu=\mu(I;{{\rho_0}})$. With this choice the state 
$\Gamma_{\rho_0,\mu}^{I}$ is separable. Since ${\cal P}$ is LOCC~\cite{dist},  also  $\Gamma_{\openone/d,\mu}^{I}$ is separable.
Therefore we must have $\mu(I;{{\openone/d}}) \leq \mu$ concluding the derivation. 
\newline 

\subsubsection{Convexity rules}\label{convexity}

Consider $\{ p_j; \Phi_j\}$ a statistical ensemble of   CPT channels $\Phi_j$ distributed according to the probabilities $p_j$.
Then  given  $\mu_j :=\mu_c(\Phi_j)$ the value assumed by the functional $\mu_c$  on the $j$-th element of the ensemble, 
it results that
\begin{equation}\label{CONVEX1}
\min_j \mu_j \leq \mu_c\left(\sum_{j} p_j \Phi_j \right) \leq \bar{\mu} \;,
\end{equation}
where
\begin{eqnarray}\label{CONVEX1max}
 \bar{\mu}=\frac{\sum_j (\mu_j p_j)/(1-\mu_j)}{\sum_k p_k/(1-\mu_k)} \leq \max_j  \mu_j \;.
 \end{eqnarray}
The proof of the first inequality trivially follows from the fact that convex combinations of separable states are still separable. To verify the second one instead 
it is convenient to introduce the probabilities 
$q_j := (\tfrac{p_j}{1-\mu_j}) /( \sum_k\tfrac{p_k}{1-\mu_k})$  and the states $\rho_j$ associated to the optimal depolarizing channel which saturates the minimization~(\ref{eq:muc}) 
for the map $\Phi_j$. Then the result simply follows by noticing that the density matrix 
\begin{equation}
\sum_j q_j  \Gamma^{\Phi_j}_{\rho_j,\mu_j}= (1-\bar{\mu}) (\Phi \otimes I)[ \psi_{+}]+\bar{\mu}
 \left(\rho_0 \otimes {\openone}/{d} \right )  
\end{equation} 
is separable (here $\rho_0$ is the state $\sum_j 
 \frac{q_j \mu_j}{ \bar{\mu}}  \rho_j $).
 
 Equations~(\ref{CONVEX1}) and~(\ref{CONVEX1max}) state that the value of $\mu_c$ associated with the average map $\sum_j p_j \Phi_j$ of $\{ p_j; \Phi_j\}$
  always lays between  the extremal values it takes on  the elements of the ensemble.

\subsection{Noise addition via concatenation}\label{sec:conc}

As evident from Sec.~\ref{moncon}
a simple method for increasing the noise level of a given map $\Phi$ is via concatenation. 
 In particular Eq.~(\ref{prop3}) implies that 
\begin{equation}\label{eq:mucn}
\mu_c(\Phi^n)\leq\mu_c(\Phi^{n-1}) \leq \ldots \leq \mu_c(\Phi^2) \leq\mu_c(\Phi)\; ,
\end{equation}
where  for each $n$ integer we have set
 \begin{eqnarray}
 \Phi^{n}:= \underbrace{\Phi\circ \Phi \circ \cdots \circ \Phi}_{n \;\; \mbox{{\small times}} } \;. \label{CONC}
 \end{eqnarray}
This allows us to introduce a new criterion for classifying noisy quantum channels. 
For this purpose,   given  $n$ integer
 we define   $\EB^n$ to be the set formed by the CPT maps $\Phi$ acting on $S$ such that 
$\Phi^n$ is entanglement-breaking, i.e. explicitly
\begin{eqnarray}
\EB^n : = \{ \Phi \in \mbox{CPT,  s. t. $\mu_c(\Phi^n) = 0$}\}\;.
\end{eqnarray}  
[Note: The set $\EB^2$ was introduced in \cite{filippov} in connection to the set of $2$-local entanglement-annihilating channels].
 Form Eq.~(\ref{eq:mucn}) it follows that 
  for all $m\geq n$ we have 
\begin{equation}\label{eq:EBn}
\EB^{n}  \subset \EB^{m}\; ,
\end{equation}
the inclusion being strict (explicit examples of maps belonging to $\EB^{m}$ but not to $\EB^{n}$ will be provided in the next sections).
Clearly a channel $\Phi$ which is an element of $\EB^{n+1}$ but which does not belong  to $\EB^n$ can be considered  ``less noisy"  than the elements of the latter set  
 as they break the entanglement shared between the system and an arbitrary ancilla just after $n$ reiterations. 
 This motivates the introduction of a new functional  $n_c(\Phi)$ to ``measure" the noise level of $\Phi$.
 In 
 particular if $\Phi\in  \EB$ we set $n_c(\Phi)=1$. Vice-versa, if for an integer number $n\geq 1$ we have 
 $\Phi\in  \EB^n/\EB^{n-1}$, we say that the map $\Phi$ is an entanglement-breaking channel of order $n$, and we
 set $n_c(\Phi)=n$. Finally if for all $n$ integers, $\Phi$ does not belong to $\EB^n/\EB^{n-1}$, then we set $n_c(\Phi)=\infty$. Trivial examples of channels with  $n_c(\Phi)=\infty$ are the unitary transformations. Other maps which do not become entanglement-breaking after any number of iterations are the amplitude-damping channels -- see Sec.~\ref{sec:ame} for details.

\subsubsection{Amendable channels}\label{sec:amendableChannels}  \label{amendableorder}
As anticipated in Sec.~\ref{sec:introduction}, differently from $\mu_c$ of Sec.~\ref{sec:convex}, 
the functional $n_c$  is not monotonic under concatenation, i.e. 
it is \emph{not} true that for an arbitrary CPT channels $\Psi$ one has  $n_c(\Psi \circ \Phi) \leq n_c(\Phi)$.  Notice however that  from Eq.~(\ref{eq:EBn}) 
it trivially follows that 
for each $\Phi \in$ CPT and  $n$ integer we still have 
\begin{eqnarray}\label{INEQ1}
n_c(\Phi^{n}) \geq n_c(\Phi^{n+1})\;.   
 \end{eqnarray}
 Furthermore exploiting the fact that the entanglement-breaking property of a map is invariant under
unitary redefinition of the input and output spaces,  and the fact that $({\cal U}\circ \Phi\circ {\cal U}^\dag)^n = {\cal U}\circ \Phi^n\circ {\cal U}^\dag$ 
we also have 
\begin{eqnarray}
 n_c({\cal U}\circ \Phi\circ {\cal U}^\dag) = n_c(\Phi)\;, \label{yesunit}
 \end{eqnarray}
${\cal U}^\dag$  being the inverse of ${\cal U}$. 

The absence of monotonicity under concatenation 
implies that $n_c$ will be typically \emph{not} invariant under unitary equivalence built upon 
uncorrelated unitary redefinitions of the input and output subspaces, i.e. 
exists $\Phi$ and  ${\cal U}$, ${\cal V}$ unitaries such that  
 \begin{eqnarray}
 \label{notunit}
  n_c({\cal U}\circ \Phi\circ {\cal V}) \neq n_c(\Phi)\;,
 \end{eqnarray}
(compare this with the identity (\ref{yesunit})).
These facts
are  strictly related with  the notion of \emph{amendable channels}. 
Specifically we say that a map $\Phi$ is amendable  
if there exists a \emph{filtering} CPT  map $\tilde{\Phi}$ such that $\Phi \circ \tilde{\Phi}\circ \Phi \notin \EB$ being $\Phi^2 \in \EB$
\cite{NOTE}. In particular we will focus on unitary filters $\tilde{\Phi}$. 
Indeed explicit examples can be found (see next section) of transformations $\Phi \in \EB^2/\EB$ for which 
there exists $\cal V$ unitary such that  $\Phi \circ {\cal V}\circ  \Phi$ (and therefore its unitarily conjugate  ${\cal V}\circ \Phi \circ {\cal V}\circ  \Phi$) is not entanglement-breaking.
Accordingly, the following inequality holds,
\begin{eqnarray} \label{ncphi}
2= n_c(\Phi) < n_c({\cal V} \circ \Phi)\;,
\end{eqnarray}    
 which explicitly disproves the monotonicity of $n_c$ under concatenation. 

The notion of amendable channels will be clarified in Sec.~\ref{sec:qubit} where, focusing on unital and generalized amplitude-damping channels for qubits, we
will  provide explicit examples of the inequality~(\ref{ncphi}). In particular in Sec.~\ref{sec:ame} we will show that, given an arbitrary integer $m\geq 2$,  it is possible to find $\Phi$ and ${\cal V}$ such that 
(\ref{ncphi}) holds with $n_c({\cal V} \circ \Phi)=m$ (maps $\Phi$ for which this is possible will be called \emph{amendable channel of order $m$}). Such examples show that there are cases where  even though two successive uses of the same channel $\Phi$ are sufficient to destroy completely the entanglement present in the system, one can \emph{delay} such detrimental effect by $m-2$ steps by simply acting with the same intermediate filtering   rotation ${\cal V}$ after each channel application. In other words, for these special maps 
it is possible to make the sequences 
 \begin{eqnarray}
\underbrace{({\cal V}\circ \Phi) \circ ({\cal V}\circ \Phi) \circ \cdots \circ({\cal V}\circ \Phi)}_{m' \;\; \mbox{{\small times}} } \;, \label{CONCm}
 \end{eqnarray}
not entanglement-breaking for all $m'<m$. 

\subsubsection{Regularization of $n_c$} \label{sec:reg}  Property~(\ref{notunit}) 
implies that the  functional  $n_c$ cannot be considered a proper \emph{measure of noise} for quantum maps but rather a \emph{criterion} for classifying them.  More stable versions of $n_c$   
can  be obtained by optimizing with respect to all filtering transformations. For instance a possibility is to
consider the regularization  
\begin{eqnarray} 
\bar{n}_c(\Phi) := \max_{\Psi}  n_c(\Psi\circ \Phi) \;,\label{reg}
\end{eqnarray} 
where the maximization is performed over all CPT maps~$\Psi$. By construction~(\ref{reg})  is 
explicitly monotonic under concatenation, i.e. 
\begin{eqnarray}
\bar{n}_c(\Phi_2\circ \Phi_1) &\leq& \bar{n}_c(\Phi_1) \;,
\end{eqnarray}  
for all  $\Phi_1$, $\Phi_2$ CPT.
Notice however that it is  still possible that \emph{not uniform} filtering CPT transformations ${\cal V}_1, {\cal V}_2,\cdots, {\cal V}_m$ might  exist which make the map 
 \begin{eqnarray}
{({\cal V}_m\circ \Phi) \circ ({\cal V}_{m-1} \circ \Phi) \circ \cdots \circ({\cal V}_1\circ \Phi)}\;, \label{CONCm1}
 \end{eqnarray}
not EB for $m\geq \bar{n}_c(\Phi)$. Removing this last instability is possible (e.g. adopting the
\emph{quantum comb} formalism of Ref.~\cite{GIULIO})
but clearly the resulting expression 
becomes extremely complex. Furthermore 
any optimization with respect to the filtering transformations tends to hide some of the structural properties of $\Phi$. For this reason, in what follows we will
restrict our attention to compute the non regularized version of $n_c$ introduced in Sec.~\ref{sec:conc}.

 \section{Qubit channels}\label{sec:qubit}
In what follows we analyze the behavior of the noise-quantifiers introduced in the previous sections by focusing on
 two important classes of qubit channels which, due to their special symmetries, allow us to  provide closed expressions for both $\mu_c$ and $n_c$.
We start in Sec.~\ref{sec:unital} by studying  the  set of qubit unital maps~\cite{ruskai}. Here  
 we will construct explicit examples of amendable channels of order 2  and prove that the sets $\EB^n$ are not convex for $n\geq 2$.
In  Sec.~\ref{sec:GAD} instead we will focus on generalized amplitude-damping channels which will allow us to show that exist 
 examples of amendable channels of arbitrary order.

\subsection{Unital channels}\label{sec:unital}
We recall that unital CPT channels transform the identity operator into itself, i.e. 
  $\Phi[\openone]=\openone$.
  This set includes unitary transformations as a proper subset, and it is closed under
  closed convex combination and channel multiplication~\cite{REV}. 
  When operating on a single qubit,  unital channels admit a simple one-to-one 
parameterization \begin{eqnarray}
\Phi \leftrightarrow T\;,  \label{correspondence}
\end{eqnarray}
 in terms of
the  $3\times3$ real matrices  $T$ which fulfills the necessary and sufficient constraint~\cite{ruskai}
 \begin{eqnarray} \label{condunital} 
  T^\dagger T\leq \openone\;,
\end{eqnarray}
(the inequality being saturated by the $T$ which correspond to unitary transformations  -- in other words unitary qubit channels 
can be uniquely associated with $3\times3$ real matrices $O$ which are orthogonal, i.e.  $O^\dag O=\openone$).
The  correspondence~(\ref{correspondence})  can be easily established by 
expanding the system density operators in the canonical operator basis 
 $\{\openone, \sigma_1,\sigma_2, \sigma_3\}$ given by the unit $\openone$ and the Pauli matrices $\sigma_i$, $1 \leq i \leq 3$. 
Accordingly, given a generic qubit density matrix $\rho=(\openone+\vec{v}\cdot\vec{\sigma})/2 \in \mathfrak{S}({\cal H}_S)$ with $\vec{v}\in \mathbb{R}^3$ such that $|\vec{v}|\leq1$, 
we can express its evolved counterpart 
$\Phi(\rho)=(\openone+\vec{v}'\cdot\vec{\sigma})/2$  via the unital CPT channel $\Phi$ as 
the linear mapping
\begin{eqnarray}
\vec{v}\rightarrow \vec{v}' = T\vec{v} \;,
\end{eqnarray}
with $T$ being $3\times3$ real matrix fulfilling~(\ref{condunital}).
It also follows that  given  two unital maps $\Phi_1$ and $\Phi_2$, characterized by the $3 \times 3$ real matrices $T_1$ and $T_2$ respectively, the composite map $\Phi_1 \circ \Phi_2$ is associated to the matrix $T=T_1 T_2$, i.e. 
\begin{eqnarray}
\label{product} 
\Phi_1 \circ \Phi_2\quad  \leftrightarrow \quad T_1 T_2 \;. 
\end{eqnarray}

 Within the correspondence~(\ref{correspondence}) it was shown \cite{ruskai} that the entanglement-breaking property of a unital qubit channel $\Phi$ only depends on the singular values of  $T$. More precisely, a unital map is EB if and only if 
\begin{equation} \label{condunitalEB1}
||T||_1=\mathrm{Tr}[\, \Lambda \,]\leq 1, \quad \Lambda:=|T|=\sqrt{T^\dagger T},
\end{equation}
where $||\cdot||_1$ is the trace norm and corresponds to the sum of the singular values of the
matrix it is applied to, while the real non-negative matrix 
$\Lambda$  is related to $T$ via polar
decomposition,
\begin{eqnarray}
T=O \Lambda \;, \label{cond111}
\end{eqnarray} 
with $O$ being
 a $3\times 3$ orthogonal matrix~\cite{HORN}.
Observe that $\Lambda$ satisfies the condition (\ref{condunital}) and  can be hence associated with
a unital qubit CPT channel via the correspondence~(\ref{correspondence}). As a matter of fact this new map $\Phi_p$ is the  \textit{polar form}  of the original channel $\Phi$ which is related to the latter through a unitary transformation according to the identity
 \begin{eqnarray} \label{ffdnew}
 \Phi=\mathcal{U}\circ \Phi_p \;,
 \end{eqnarray}
 which mimics~(\ref{cond111}) at the super-operator level via the identity~(\ref{product}) 
  (here   $\mathcal{U}$ is the unitary channel which is associated  to the orthogonal matrix $O$ of Eq.~(\ref{cond111}) through the correspondence~(\ref{correspondence})). 
The above construction in particular shows also that, without loss of generality, when characterizing the
entanglement-breaking property of a map $\Phi$ we can just focus on its polar form $\Phi_p$. At the level of super-operators this follows simply from~(\ref{ffdnew}) and by the fact that channels which differ by 
unitary transformations share the same entanglement-breacking property
(alternatively this can also
verified through Eq.~(\ref{condunitalEB1}) and noticing that 
the  singular values of $\Lambda$ coincides with those of $T$ so that  $\| \Lambda\|_1 =\|T\|_1$).

The decomposition~(\ref{cond111}) can be
further simplified by 
expressing $T$ as 
\begin{eqnarray} \label{TNEWDEC}
 T = O_1 \;{{D}} \;O_2\;,
 \end{eqnarray}
 where $O_{1,2}$ are orthogonal matrices, while ${{D}}=\mathrm{diag} \{\lambda_1,\lambda_2,\lambda_3\}$ is a diagonal (not necessarily positive) matrix with 
 elements $\lambda_j$, which in modulus correspond to the  singular values of $T$
 (i.e.  to the eigenvalues of $\Lambda$) so that 
\begin{equation}\label{eq:canonicalEB}
||T||_1 = \sum_{i=1}^3 |\lambda_i|\,.
\end{equation}
In terms of the equivalence~(\ref{correspondence}), the matrix  ${{D}}$ defines a unital channel 
$\Phi_{{\lambda}}$ which gives the  
 \textit{canonical representation}  of the map $\Phi$~\cite{ruskai}, and which is related 
 to the latter through the identity 
 \begin{eqnarray} \label{identitynew1}
 \Phi = {\cal U}_1 \circ \Phi_{{\lambda}} \circ {\cal U}_2\;,
 \end{eqnarray} 
with ${\cal U}_{1,2}$ being  unitary transformations associated to the matrix $O_{1,2}$ through
the isomorphism~(\ref{correspondence}). As in the case of the polar representation, it is clear that 
$\Phi$ (or $\Phi_p$) is entanglement-breaking
 if and only if $\Phi_{{\lambda}}$ is entanglement-breaking, i.e. 
\begin{eqnarray}
\Phi\in \EB \; \Leftrightarrow \; \Phi_p \in \EB \; \Leftrightarrow \; \Phi_{{\lambda}} \in \EB \;.
\label{equivalence}
\end{eqnarray}

\subsubsection{Noise addition via convex convolution}
For a unital qubit channel $\Phi$ described by the matrix $T$, the value $\mu_c(\Phi)$ can be exactly computed resulting in the following increasing function of $\|T\|_1$, 
\begin{eqnarray} \label{value1} 
\mu_c(\Phi) = \max\left\{ 0, \frac{ \| T\|_1 -1}{\|T\|_1}\right\} \;,
\end{eqnarray} 
which for unitary transformations  gives
$\mu_c(\Phi)=2/3$ in agreement with Eq.~(\ref{boundachi}) (in this case in fact 
$\| T\|_1=3$), while for 
 entanglement-breaking channels  it correctly gives $\mu_c(\Phi)=0$  (in this case in fact 
 $\| T\|_1\leq 1$).
 
To derive Eq.~(\ref{value1}) we first exploit the property~(\ref{eq:mucinvariance}) and the identity  (\ref{identitynew1})  to write   
\begin{equation} 
\mu_c(\Phi)=\mu_c(\Phi_p)=\mu_c(\Phil)\;.
\end{equation}
Now the value of $\mu_c(\Phil)$ can be directly computed by noticing that 
\begin{equation}\label{eq:mucUnitali}
\mu_c (\Phil)=\mu\left(\Phi_{{\lambda}}; {{\openone/2}}\right)
\;.
\end{equation}
Proving this identity requires us to show that  if there exists $\rho_0 \in \mathfrak{S}({\cal H}_S)$ such that for a given $\mu\in[0,1]$, the Choi-Jamiolkowski state~(\ref{eq:CJ})
$\Gamma^{\Phil}_{\rho_0,\mu}$ is separable, then for that same $\mu$, the state $\Gamma^{\Phil}_{\openone/2,\mu}$ is also separable. 
This follows by the fact that the canonical channels $\Phil$ satisfy the following relation (see Appendix~\ref{sec:KrausUnitalQubit}),
\begin{equation}
\Phil=\mathcal{S}^\dag_j \circ \Phil\circ  \mathcal{S}_j\;, \label{commuting}
 \end{equation}
where for  $j \in \{ 0,1,2,3\}$ the symbol ${\cal S}_j={\cal S}_j^\dag$ stands for the unitary mapping 
\begin{eqnarray} \label{superss}
 \mathcal{S}_j[\rho]:=\sigma_{j} \rho \sigma_j\;,
 \end{eqnarray} 
 (here we used  $\sigma_0 := \openone$). 
Suppose then  that for some $\rho_0$ and $\mu$ the state $\Gamma^{\Phil}_{\rho_0,\mu}$  is separable. 
This implies that also $({\cal S}_j \otimes {\cal S}_j)(\Gamma^{\Phil}_{\rho_0,\mu})$ is separable for all $j$
(indeed this is just $\Gamma^{\Phil}_{\rho_0,\mu}$ transformed  under local rotations). Notice that such state
can be expressed as 
  \begin{eqnarray}\nonumber
  &&\!\!\!\!\!\!\!({\cal S}_j \otimes {\cal S}_j)(\Gamma^{\Phil}_{\rho_0,\mu}) =
(1-\mu)( \Phil \circ  \mathcal{S}_j\otimes {\mathcal S}_j)[\psi_{+}] \\
&&\qquad \qquad \qquad \qquad \qquad \quad \;\;  +\; \mu\big(\mathcal{S}_j[\rho_0]\otimes \frac{\openone}{2}\big) \nonumber \\ 
&&  \;\;\;\;\; =   \label{eq:dep0sep}
(1-\mu)( \Phil \otimes I)[\psi_{+}] +\; \mu\big(\mathcal{S}_j[\rho_0]\otimes \frac{\openone}{2}\big)\;,
\end{eqnarray}
where in the first line  we used Eq.~(\ref{commuting}) 
and the fact that $ {\mathcal S}_j\circ  {\mathcal S}_j=I$, while in the last one we used the fact that 
$(\mathcal{S}_j\otimes {\mathcal S}_j)[\psi_+]=\psi_+$ \cite{NOTA}. 
Next we observe  that $\Gamma^{\Phil}_{\openone/2,\mu}$
can be obtained from (\ref{eq:dep0sep}) by averaging over $j$ (indeed $\sum_{j=0} ^ 3\mathcal{S}_j /4=\Phi^{\openone/2}_{\DEP}$ and hence $\sum_{j=0} ^ 3\mathcal{S}_j[\rho_0] /4={\openone/2}$). 
Therefore, since mixtures of separable states are separable, we conclude that $\Gamma^{\Phil}_{\openone/2,\mu}$ is separable, proving Eq.~(\ref{eq:mucUnitali}). The final step in  deriving Eq.~(\ref{value1})
is  hence the evaluation of $\mu\left(\Phi_{{\lambda}}; {{\openone/2}}\right)$. This however is
easily obtained by noticing that $(1-\mu)\Phi_{{\lambda}}+\mu\Phi^{{\openone}/{2}}_\DEP$ is a unital
channel which at the level of the correspondence~(\ref{correspondence}) 
is characterized by a matrix $T'$ having
singular values $|\lambda'_j|=  (1-\mu) |\lambda_i|$. Therefore according to 
Eq.~(\ref{condunitalEB1})  it is entanglement-breaking for
\begin{equation}
 \sum_{i=1}^3 (1-\mu) |\lambda_i|\leq1 \Longleftrightarrow  \mu \geq \frac{\sum_{i=1}^3 |\lambda_i|-1}{\sum_{j=1}^{3} |\lambda_i|}=\frac{ \| T\|_1 -1}{\|T\|_1}\;,
\end{equation}
yielding 
\begin{eqnarray} \label{value2} 
\mu\left(\Phi_{{\lambda}}; {{\openone/2}}\right) = \max\left\{ 0, \frac{ \| T\|_1 -1}{\|T\|_1}\right\} \;,
\end{eqnarray} 
and hence concluding the calculation. 

\subsubsection{Noise addition via concatenation and amendable unital channels} \label{secIII2A}
For what concerns the mechanism of noise addition via concatenation, we start noting that
differently from the previous section, 
 a unital qubit channel $\Phi$ and its polar $\Phi_p$ or canonical  $\Phi_\lambda$ forms  might 
not share the same properties: indeed, as anticipated in Sec.~\ref{sec:amendableChannels}, the functional $n_c$ is not invariant under unitary equivalence (an explicit counterexample will be provided in the following).
Still it turns out that $\Phi_p$ and $\Phi_\lambda$ have same the value of $n_c$~\cite{NOTA3}
and that  a sufficient (but not necessary) condition for a generic unital channel $\Phi$  to be EB$^n$ is $\Phi_p \in \EB^n$, implying 
\begin{equation} \label{ineq22}
n_c(\Phi_p) \geq n_c ( \Phi)=n_c (\mathcal{U}\circ \Phi_p)\,.
\end{equation}
To prove this relation we first notice that from Eqs.~(\ref{product}) and (\ref{condunitalEB1}) it follows that
$\Phi^n$ is entanglement-breaking if and only if  
\begin{equation} \label{condunitalEB}
||T^n||_1\leq 1\;,
\end{equation}
where $T$ is the matrix associated to $\Phi$ via the correspondence~(\ref{correspondence}).
Hence, reminding Eq.~(\ref{INEQ1}) we can write 
\begin{eqnarray} \label{COMPUTENC}
n_c(\Phi) = \min_{n} \{n\;\; \mbox{s.t.} \;\; \|T^n\|_1\leq 1\} \;,
\end{eqnarray} 
where the minimization is performed over all $n$ integers (notice that for entanglement-breaking channels 
this yields correctly $n_c(\Phi)=1$).
Equation~(\ref{ineq22}) can then be derived by noticing that the matrices $T$ and $\Lambda$ which represent $\Phi$ and $\Phi_p$, respectively, are related as in (\ref{cond111}) and by using 
the H\"older inequality~\cite{HOLD}. Indeed 
\begin{eqnarray}
\!\!||T^n||_1 = ||(O \Lambda)^n||_1 &\leq& (||O \Lambda||_n)^n= (||\Lambda||_n)^n\nonumber\\
&=&\left(\mathrm{Tr}[\Lambda^n]^{1/n}\right)^n=||\Lambda^n||_1\;, 
\end{eqnarray}
where $\| \cdot \|_n$ is the operator $n$-norm, and where we used the fact that $O$ is orthogonal and $\Lambda$ non-negative. From this we can hence  conclude that if $n_c(\Phi_p)=n$ (that is $||\Lambda^n||_1\leq 1$), then $\Phi=\mathcal{U}\circ\Phi_p \in$ EB$^n$ too ($||(O\Lambda)^n||_1\leq 1$), concluding the proof.

On the contrary, for $n\geq 2$ there exist $\Phi\in \EB^n$ such that their polar form $\Phi_p$ is not in  EB$^n$~\cite{NOTA2}. As an example consider for instance the case in which $\Phi_p$ is characterized by the matrix 
\begin{eqnarray} \label{example} 
\Lambda=\left[ \begin{array}{ccc}
0.73 & 0 &0 \\
0 &0.5 &0 \\
0&0& 0.5 
\end{array}
\right]\;,
\end{eqnarray}
which satisfies the condition~(\ref{condunital}) and for which one has 
\begin{eqnarray} \label{fffddd}
&&||\Lambda||_1 =1.73>1\;, \qquad   ||\Lambda^2||_1\simeq1.03>1 \;,  \nonumber \\
&& ||\Lambda^3||_1\simeq0.64 <1\;,
\end{eqnarray} 
that imply   $n_c(\Phi_p)=3$ and  hence $\Phi_p\in \EB^3/\EB^2 \subset \EB^3$. Now construct $\Phi$ 
as the unital map associated with the matrix $T$ defined as in Eq.~(\ref{cond111})  where $O$ is the  orthogonal matrix 
\begin{eqnarray} \label{orto} 
O=O^\dag=\left[ \begin{array}{ccc}
0 & 1 &0 \\
1 &0 &0 \\
0&0& 1 
\end{array}
\right]\;,
\end{eqnarray}
that is 
\begin{eqnarray} \label{newT} 
T = O \Lambda=\left[ \begin{array}{ccc}
0 & 0.5 &0 \\
0.73 &0 &0 \\
0&0& 0.5 
\end{array}
\right]\;.
\end{eqnarray}
With this choice we have 
\begin{eqnarray}
\| T^2\|_1 =0 .98 <1\;,
\end{eqnarray}
which gives   $n_c(\Phi)=2$ and thus $\Phi \in \EB^2/\EB\subset \EB^2$.

The same example proves the existence of unital qubit maps that are amendable. To see this explicitly take ${\cal V}$ of Eq.~(\ref{ncphi}) as 
the inverse ${\cal U}^\dag$ of the unitary ${\cal U}$ of Eq.~(\ref{ffdnew}) and notice that 
while the map $\Phi$ introduced above  is an element of  $\EB^2$ (i.e. $\Phi^2 \in \EB$),  the channel 
$\Phi \circ {\cal U}^\dag \circ \Phi =  \mathcal{U}\circ\Phi_p^2$ is not entanglement-breaking (it is indeed unitarily equivalent to $\Phi_p^2$ which is not in $\EB$ according to Eq.~(\ref{fffddd})):
this implies that, by interposing a unitary transformation, we can prevent two consecutive applications of $\Phi$ of becoming entanglement-breaking. 
For the sake of completeness  is worth mentioning that this fact is in contradiction with one of the  claims of Ref.~\cite{gavenda} where it was stated (without presenting a formal proof) that unital qubit channels are 
no rectifiable (and hence no amendable). At the origin of this incompatibility between such claim and our findings, we believe is the fact that  
 the Authors of~\cite{gavenda}  assumed the possibility of always putting the channel in the canonical representation~(\ref{identitynew1}),
without considering the fact that the properties of noise addition via concatenation are \emph{not} invariant under unitary equivalence.

  \subsubsection{Non convexity of the sets $\EB^n$} 
The set of $\EB$ channels is known to be convex. The same property however does not apply for the set  $\EB^n$ with $n\geq 2$. 
In what follows we will show this fact by 
presenting counter-examples taken from the unital qubit maps for $n=2, 3$
(even though we did not
check for larger $n$ we believe that the similar counter-examples can be found also for those cases).
  Specifically we will prove  that 
even though convex combinations of untial channels are still unital, the  unital qubit channels which are in EB$^2$ and in $\EB^3$ are not closed under convex convolution.
 
\paragraph{Non convexity of $\EB^2$:}  \label{paraa}  Via the correspondence (\ref{correspondence}) the fact  that convex combinations of qubit unital channels in $\EB^2$ are not  necessarily elements of the same set, 
can be traced back to the  fact  that the set formed by operators $T$ satisfying the condition 
\begin{eqnarray}
\| T^2 \|_1 &\leq& 1 \;\label{PROffP122}
\end{eqnarray}    
is not convex. To show this thesis it is sufficient to provide a counter-example. 
For instance take $T$ as in (\ref{newT}) and its transpose $T^\top$. Since both these matrices satisfy the condition~(\ref{condunital}) 
they  properly define two unital qubit channels $\Phi$ and $\Psi$ via the correspondence~(\ref{correspondence}). Furthermore
since 
\begin{eqnarray} \label{condt2}
\| T^2\|_1 = \|(T^\top)^2\|_1 =0 .98 <1\;,
\end{eqnarray}
these maps  belong to $\EB^2$. Take then the channel $\bar{\Phi} = (\Phi + \Psi)/2$: it is clearly unital and it is described by a matrix~(\ref{correspondence}) 
formed by the convex convolution 
\begin{eqnarray}
\bar{T} = (T + T^\top )/2 = 
\left[ \begin{array}{ccc}
0 & 0.615&0 \\
0.615  &0 &0 \\
0&0& 0.5 
\end{array}
\right]\;. \label{tmedio}
\end{eqnarray}
Computing its $\| \bar{T}^2\|_1$ we notice however that it does not fulfill~(\ref{condt2}), indeed 
\begin{eqnarray} 
\| \bar{T}^2\|_1\simeq1.01
>1 \;, 
\end{eqnarray}
implying that $\bar{\Phi}$ is not in $\EB^2$, 
completing hence the proof.

\paragraph{Non convexity of $\EB^3$:} \label{parab} 

A counter-example can be obtained by taking the unital channels $\Phi$ and $\Psi$, and $\bar{\Phi} = (\Phi + \Psi)/2$ 
associated with matrices $T$, $T^\top$, and $\bar{T}= (T+T^\top)/2$ where now however  $T$ is expressed as $O\Lambda$
with $O$ is agin as in Eq.~(\ref{orto}) but with $\Lambda$ being 
\begin{eqnarray}
\Lambda=\left[ \begin{array}{ccc}
0.91 & 0 &0 \\
0 &0.6 &0 \\
0&0&0 .55 
\end{array}
\right]\;.
\end{eqnarray}
With this choice in fact we have  $\| T^3\|_1 = \|(T^\top)^3\|_1 \simeq 0.991 <1$ (hence $\Phi$ and $\Psi$ are in $\EB^3$) but 
$\| \bar{T}^3\|_1\simeq1.03 >1$ which instead implies that $\bar{\Phi} \notin \EB^3$. 
\newline 

The example~(\ref{tmedio}) seems to contradict one of the claims of~\cite{filippov}  which states that the set of unital qubit 
 $\EB^2$ is closed under convex convolution. As for the incompatibility with~\cite{gavenda} underlined at the end of Sec.~\ref{secIII2A}, this 
  originates from the fact that Ref.~\cite{filippov} always assumes the canonical structure~(\ref{identitynew1}) when analyzing the untial qubit channels. 
Indeed, even though, in general the set of unital qubit channels which are $\EB^n$ is not convex,
 it is true that the subset formed by  qubit unital maps of $\EB^n$ which are in polar (or canonical) forms is convex. For $n=2$ this was explicitly  proved in \cite{filippov}. Here we generalize this fact to arbitrary $n$.
  Let then $\Phi_{p_1}$ and $\Phi_{p_2}$ be unital qubit maps in polar form  which are EB$^n$, that is $||{\Lambda_1}^n||_1\leq1$ and $||{\Lambda_2}^n||_1\leq1$
  ($\Lambda_1$ and $\Lambda_2$ being the non-negative  matrices that define the two maps within the correspondence~(\ref{correspondence})).  
According to the H\"older inequality~\cite{HOLD} we have that given an arbitrary product ${\Lambda_{i_1}}{\Lambda_{i_2}}\ldots{\Lambda_{i_n}}$ with 
  $\{i_{\alpha}\}_{1\leq\alpha \leq n}=1,2$, one has 
\begin{eqnarray}
||{\Lambda_{i_1}}{\Lambda_{i_2}}\ldots{\Lambda_{i_n}}||_1 &\leq& ||{\Lambda_{i_1}}||_{n}||{\Lambda_{i_2}}||_{n}\ldots ||{\Lambda_{i_n}}||_n \nonumber\\&=&\mathrm{Tr}[{\Lambda_{i_1}}^{n}]^{\frac{1}{n}} \mathrm{Tr}[{\Lambda_{i_2}}^{n}]^{\frac{1}{n}}\ldots\mathrm{Tr}[{\Lambda_{i_n}}^{n}]^{\frac{1}{n}}\nonumber\\&=&\Big(||{\Lambda_{i_1}}^n||_1 ||{\Lambda_{i_2}}^n||_1\ldots ||{\Lambda_{i_n}}^n||_1\Big)^{1/n}\nonumber\\&\leq& 1 \,.\end{eqnarray}
Thus, exploiting the sub-additivity of the trace norm, we get that for arbitrary  probabilities $q_1, q_2$ we have 
\begin{eqnarray}
||(q_1  \Lambda_1 + q_2 \Lambda_2 )^n||_1&\leq& \sum_{i_1=1}^{2} \ldots \sum_{i_n=1}^{2} q_{i_1}\ldots q_{i_n} ||{\Lambda_{i_1}}   \ldots  \Lambda_{i_n}||_1\nonumber\\&\leq& \left(q_1+q_2\right)^n=1\,,
\end{eqnarray} 
which proves that the mixed channel $q_1 \Phi_{p_1} + q_2 \Phi_{p_2}$ is an element of $\EB^n$. 
The convexity of the set of unital qubit maps in canonical form can be proved in an similar way.

\subsubsection{Geometry of unital EB$^n$ maps in canonical form}

\begin{figure}[t]
\begin{center}
\includegraphics[width=1\columnwidth]{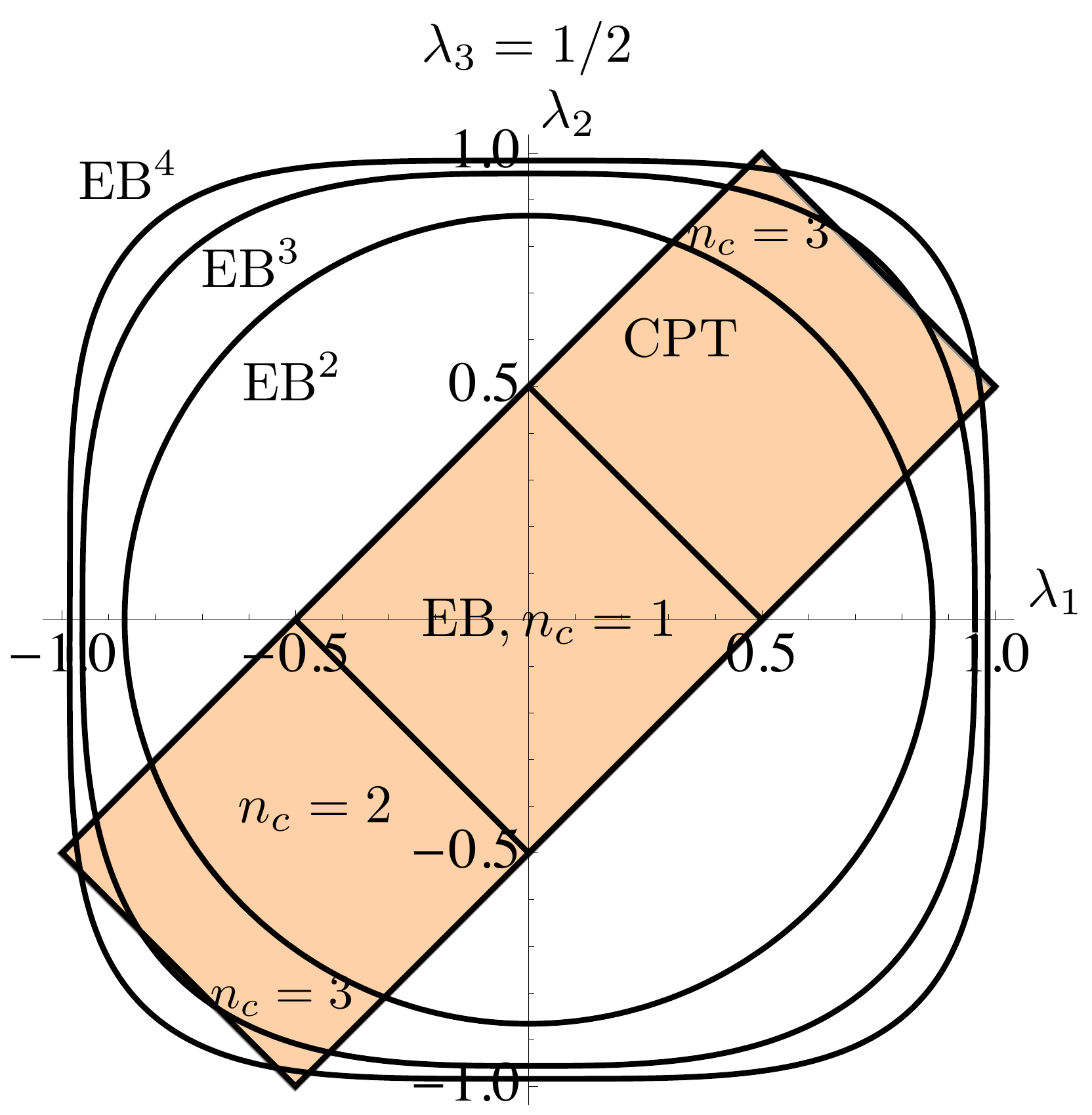}
\caption{(Color online) Section of the regions~(\ref{regionn}) which, for   qubit maps in canonical form~(\ref{identitynew1}),
identify  the entanglement-breaking qubit maps of order $n$ (here we set $\lambda_3=1/2$). The shaded region (orange in the online version) refers to the tetrahedron~(\ref{tetra}) 
which identifies all CPT maps $\Phi_\lambda$. The domain associated to the EB set  is the square at the center of the figure and corresponds to the section of the octahedron~\cite{ruskai}. 
It is the only region~(\ref{regionn}) which is  completely contained in the tetrahedron. The circle of radius $\sqrt{3}/2$ identifies instead the condition for $\Phi_\lambda$ to be $\EB^2$.}\label{fig:EBKunitali}
\end{center}
\end{figure}

For unital maps in canonical form~(\ref{identitynew1}), the necessary and sufficient condition~(\ref{condunital})  for the complete positivity can be expressed as~\cite{ruskai}
\begin{equation}
|\lambda_1 \pm \lambda_2|\leq|1\pm\lambda_3|\;, \label{tetra}
\end{equation}
which, within the parameter space of the real 3-D vectors $(\lambda_1, \lambda_2, \lambda_3)$, identifies  a  tetrahedron characterized by the extreme points $(1,1,1), (1,-1,-1), (-1,1,-1), (-1,-1,1)$. Similarly from Eqs.~(\ref{condunitalEB1}) and~(\ref{eq:canonicalEB}) we have that the set of unital entanglement-breaking maps $\EB$ for qubits in canonical form is restricted to the octahedron  of vertices $(\pm1,0,0)$ and permutations, which  correspond to the intersections between the tetrahedron  and its inversion through the origin \cite{ruskai}. The generalization to the case  EB$^n$ is straightforward. Indeed from Eqs.~(\ref{TNEWDEC}) and~(\ref{condunitalEB}) it immediately follows that  the canonical forms $\Phi_\lambda$ which are $\EB^n$ are identified by the intersection between the tetrahedron and the
  region
  \begin{eqnarray} \label{regionn}
  \sum_{0\leq i \leq 3} |\lambda_i|^n\leq 1\;,
   \end{eqnarray}
(for instance, the set $\EB^2$ corresponds to the intersection between the tetrahedron and the unit  ball  in the origin, see also Fig.~2 of Ref.~\cite{filippov}). For sake of simplicity, in Fig.~\ref{fig:EBKunitali} we have reduced the tridimensional $(\lambda_1, \lambda_2, \lambda_3)$'s space to the space $(\lambda_1,\lambda_2 )$ by fixing $\lambda_3=1/2$, and studied the domain associated to the sets $\EB^n$, $n\geq1$.
The shaded region (orange in online version), refers  to the untial maps -- i.e. it is a section of the tetrahedron~(\ref{tetra}). 
Notice that while the EB square is  contained in the CPT set, this in not the case for higher order entanglement-breaking maps. The corners of the CPT rectangle, $\lambda_1=\pm \lambda_3$ with $\lambda_2=\pm1$ and $\lambda_2=\pm \lambda_3$ with $\lambda_1=\pm1$, refer to unital maps which can never belong to EB$^n$ for finite $\lambda_3$, as the $\EB^n$ condition reduces to $2 |\lambda_3|^n\leq 0$, $n\geq1$.  Finally, observe that $\EB^{n}  \subseteq \EB^{n+1}, n \geq 1$ (see Eq.~(\ref{eq:EBn})).

\subsection{Generalized amplitude-damping channels}\label{sec:GAD}

A generalized  qubit amplitude-damping channel $\Psi_{p,\gamma}$ describes the physical process through which the system  approaches  equilibrium with its environment by spontaneous emission at finite temperature~\cite{nielsen}. This is done by modifying the statistics of the populations associated to the ground state $|0\rangle$ and the excited state $|1\rangle$ which, in the following we identify with the
eigenvectors of the Pauli matrix $\sigma_z$. We can represent them as 
\begin{equation}\label{thisclass}
\Psi_{p,\gamma}[\rho]=\sum_{1\leq i \leq 4} E_i \rho E_i^\dagger,
\end{equation}
where $E_i$ are the Kraus operators \cite{KRAUS} which, in the canonical basis $\{|0\rangle,|1\rangle\}$, are expressed by the matrices 
\begin{eqnarray}\label{defE}
E_1&=\sqrt{\gamma}\begin{pmatrix}1&0\\0&\sqrt{1-p}\end{pmatrix}, \quad E_2=\sqrt{\gamma}\begin{pmatrix}0&\sqrt{p}\\0&0\end{pmatrix}\nonumber\\ \nonumber\\
E_3&=\sqrt{1-\gamma}\begin{pmatrix}\sqrt{1-p}&0\\0&1\end{pmatrix}, \quad E_4=\sqrt{1-\gamma}\begin{pmatrix}0&0\\\sqrt{p}&0\end{pmatrix},\nonumber\\
\end{eqnarray}
with $0 \leq p \leq 1$ and $0 \leq\gamma \leq 1$.

For these channels we will provide an explicit expression for the functional $\mu_c(\Psi_{p,\gamma})$ and compare it with  $\mu_c(\Psi^2_{p,\gamma})$ verifying Eq.~(\ref{eq:mucn}). 
We will also compute explicitly the value of    $n_c(\Psi_{p,\gamma})$ and construct examples of  amendable channels of arbitrary order. Before entering in the details of these calculations however we
observe that from Eq.~(\ref{defE}) it follows
\begin{eqnarray}
\Psi_{p,1-\gamma} = {\cal S}_1 \circ \Psi_{p,\gamma} \circ {\cal S}_1^\dag\;,
\end{eqnarray} 
where  ${\cal S}_1$ is the Pauli rotation~(\ref{superss}). This implies that  we can restrict the analysis to the case $\gamma\in[0,1/2]$ since the following identities hold
\begin{eqnarray}
\mu_c( \Psi^n_{p,1-\gamma} ) &=& \mu_c( \Psi^n_{p,\gamma} ) \;, \qquad \forall n \;, \\
n_c( \Psi_{p,1-\gamma} ) &=& n_c( \Psi_{p,\gamma} ) \;, \label{ncnewid}
\end{eqnarray} 
(the first deriving from Eq.~(\ref{prop5}) while the second from~(\ref{yesunit})).

\subsubsection{Noise addition via convex convolution}
By definition, in order to determine $\mu_c$  we have to perform two minimizations: the first  leads to  $\mu(\Psi_{p,\gamma}; {{\rho_0}})$, and the second  consists of a minimization over all possible choices of $\rho_0$ (see Eqs.~(\ref{eq:semimuc}) and~(\ref{eq:muc})). 
Regarding the second step it is useful to observe that for the maps~(\ref{thisclass}) we can focus on $\rho_0$ lying on the $z$ axis, i.e. on states with the following Bloch   form
\begin{eqnarray} \label{lungoz}
 \rho_0^{z}=\frac{\openone+ v_z \sigma_z }{2}\,.
\end{eqnarray}
This can be proved following the same line of reasoning  that led us to the identy~(\ref{eq:mucUnitali}) via  Eq.~(\ref{commuting}). Indeed we notice that the channel $\Phipg$ commutes with 
 all the  unitary rotations ${U}_z=\exp[i \theta \sigma_z]$ around the $z$ axis, i.e. 
\begin{eqnarray}
  \Phipg  =\mathcal{U}_z^\dag \circ  \Phipg \circ \mathcal{U}_z\;,
\end{eqnarray} 
being $\mathcal{U}_z[\rho]:= U_z \rho U_z^\dagger$.
We can thus conclude that if 
there is a state  $\rho_0=({\openone + \vec{v}\cdot \vec{\sigma}})/{2}$ with arbitrary $\vec{v}\in \mathbb{R}^3$  such that $\Gamma_{\rho_0,\mu}^{\Phipg}$ is separable for some $\mu$, then also the state $\int dm({U}_z) (\mathcal{U}_z \otimes \mathcal{U}_z)[\Gamma_{\rho_0,\mu}^{\Phipg}]$
obtained by mixing over all the local rotations ${\cal U}_z\otimes {\cal U}_z$ will be separable ($dm({U}_z)$ being the measure over the set of unitary rotations ${U}_z$). The latter however 
is given by 
\begin{eqnarray}
&&\int dm({U}_z) (\mathcal{U}_z \otimes \mathcal{U}_z)[\Gamma_{\rho_0,\mu}^{\Phipg}]  =(1-\mu)(\Phipg \otimes I)[\psi_{+}]\nonumber\\&&\qquad \qquad \qquad\qquad  +\mu\left(\int dm({U}_z)\mathcal{U}_z [\rho_0] \otimes \frac{\openone}{2}\right) \;,
\end{eqnarray}
which coincides with $\Gamma_{\rho_0^z,\mu}^{\Phipg}$ with $\rho_0^z$ being the density matrix of the form~(\ref{lungoz}) defined by 
\begin{equation}
\int dm ({U}_z)\; \mathcal{U}_z [\rho_0]  = \rho_0^z\;.
\end{equation}
Thus if $\Gamma_{\rho_0,\mu}^{\Phipg} \in$ SEP, then $\Gamma_{\rho^z_0,\mu}^{\Phipg} \in$ SEP which implies $\mu(\Phi; {{\rho_0^z}}) \leq \mu(\Phi; {{\rho_0}})$ as anticipated.

To determine $\mu(\Phipg; {{\rho_0^z}})$,  we recall that for  two qubits  the state  $\Gamma_{\rho_0^z,\mu}^{\Phipg}$ is separable iff, when partially transposed, it shows a positive determinant \cite{augusiak}. Exploiting this we can then write
\begin{widetext}
\begin{eqnarray}
\mu(\Phipg; {{\rho_0^z}}) = 
\label{minimize}
\frac{p \left[4 p (\gamma -1) \gamma -2 \gamma  v_z+v_z-3\right]+4-\sqrt{p^2 \left(v_z-2 \gamma +1\right){}^2+4 p \left(v_z^2-1\right)-4 v_z^2+4}}{4 p^2 (\gamma -1) \gamma +2 p (-2 v_z\gamma +v_z-1)+v_z^2+3},
\end{eqnarray}
\end{widetext} 
which needs then to be minimized over  $v_z \in [-1,1]$. Remembering that $\gamma \leq 1/2$,  we find that for 
\begin{eqnarray}
p &\leq &\bar{p}  := \frac{\sqrt{4 \gamma ^2-8 \gamma +5}-1}{2 (1-\gamma )^2}\;,
\end{eqnarray} 
the function~(\ref{minimize}) has a global minimum at $v_z =\bar{v}_z \in ]-1,1[$ 
\begin{equation}
\bar{v}_z (\gamma,p)=\frac{p \left(p+2 \sqrt{1-p}\right) (1-2 \gamma )}{4-p (p+4)}\,,
\end{equation}
yielding 
\begin{equation} \label{down}
\mu_c( \Psi_{p\leq \bar{p},\gamma\leq 1/2})
= \frac{p^2+3 p+2 \sqrt{1-p}-4}{p^2+2 p-3}\,,
\end{equation}
(notice that in this case, $\mu_c( \Psi_{p,\gamma})$ does not depend on $\gamma$).
Vice-versa, for
$p \geq \bar{p}$
the quantity  $\mu(\Phipg; {{\rho_0^z}})$ is a monotonous decreasing function of $v_z$, whose minimum corresponds to $v_z=1$ and is given by
\begin{equation} \label{up}
\mu_c(\Psi_{p\geq \bar{p},\gamma\leq 1/2}) = \frac{p [p (\gamma -1) \gamma -1]+1}{p \gamma  [p (\gamma -1)-1]+1}\;.
\end{equation}
The function $\mu_c(\Phipg)$ is continuous at $p=\bar{p}$ where both~(\ref{down}) and~(\ref{up}) 
yield
\begin{equation}
\mu_c(\Psi_{ \bar{p},\gamma\leq 1/2})=\frac{2-4 \gamma }{-4 \gamma +\sqrt{4 (\gamma -2) \gamma +5}+3}\;,
\end{equation}
which vanishes for $\gamma=1/2$. In this case we have $\bar{p}(\gamma=1/2)= 2 (\sqrt{2}-1)\simeq 0.828$ and $\Psi_{p,\gamma} \in$ EB (this point belongs to the curve $p=p_1(\gamma)$ computed in the next subsection, see Eq. (\ref{eq:pEBn})).

Similar results are found  for $\Phipg^2:=\Phipg \circ \Phipg$. More precisely we have
\begin{eqnarray}
\mu_c(\Psi^2_{p\leq \bar{\bar{p}},\gamma\leq1/2}) &=& \frac{p^2-4 p+2}{p^2-4 p+3}\\
\mu_c(\Psi^2_{p\geq \bar{\bar{p}},\gamma\leq1/2}) &=&\frac{(p-2) p [(p-2) p (\gamma -1) \gamma +1]+1}{(p-2) p \gamma  [(p-2) p (\gamma -1)+1]+1},\nonumber\\
\end{eqnarray}
with
\begin{equation}
\bar{\bar{p}}(\gamma):=\frac{\sqrt{4 \gamma ^2-8 \gamma +5}+2 \gamma -3}{2 (\gamma -1)}.
\end{equation}
Furthermore, $\mu_c(\Psi^2_{p,\gamma})$ vanishes for $\{\bar{p},\gamma\}=\{2 - \sqrt{2}, 1/2\} \simeq \{ 0.586, 1/2 \}$. We also find that incidentally the following identity holds, 
$\mu_c(\Psi_{\bar{p},\gamma})=\mu_c(\Psi^2_{\bar{\bar{p}},\gamma})$. The analysis gives analogous results for $\Phipg^n$, $n>3$. In the inset in Fig.~\ref{fig:PEBn}, as an example, we have shown the behavior of $\mu_c(\Psi_{{p},\gamma})$ and $\mu_c(\Psi^2_{{p},\gamma})$ for $\gamma=1/3$.

\subsubsection{Amendable channels of arbitrary order} \label{sec:ame}
\begin{figure*}
\begin{center}
\includegraphics[height=0.4\textwidth]{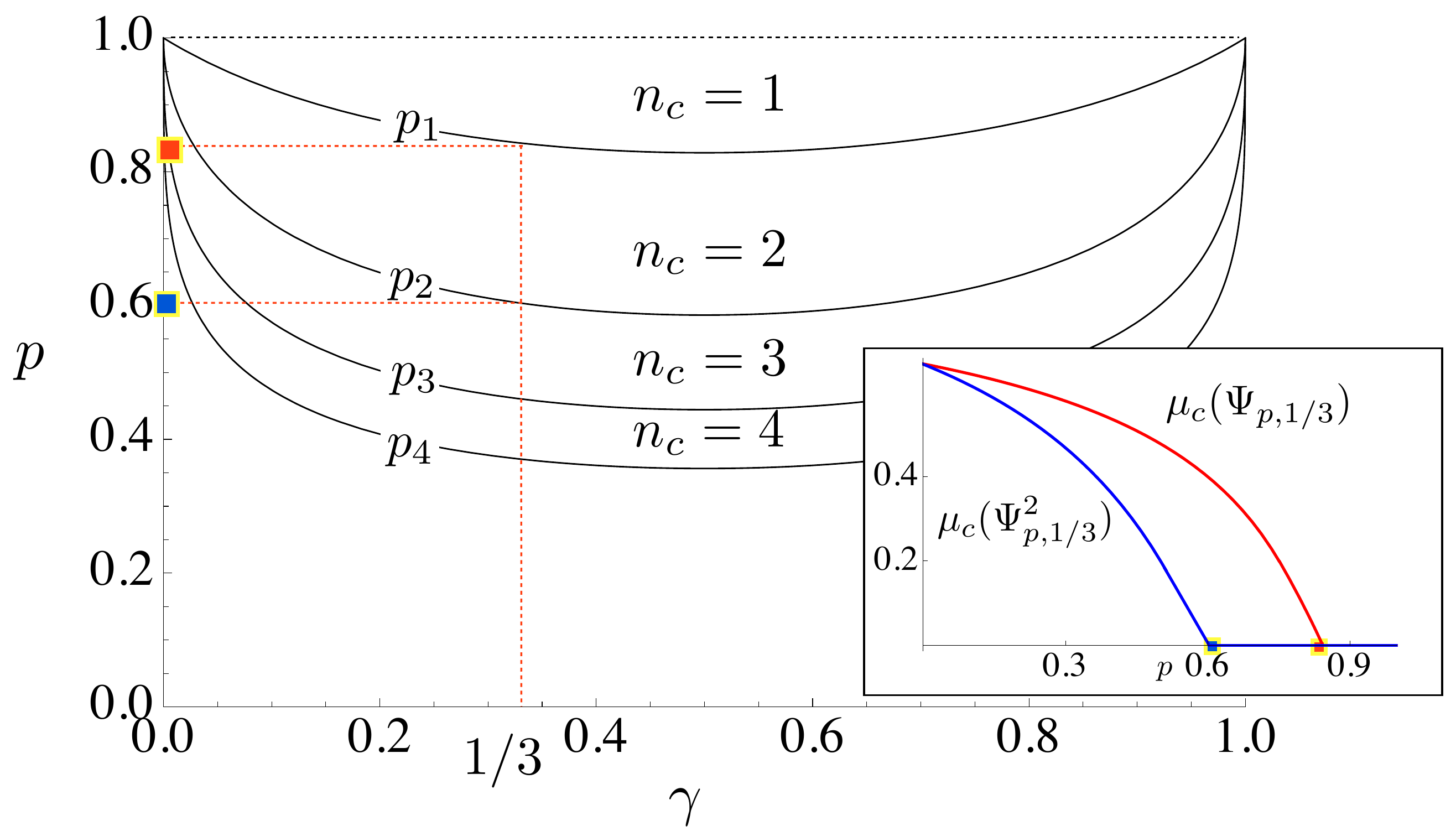}
\caption{Regions of the space $\{p,\gamma\}$ associated with channels $\Psi_{p,\gamma}$ which have $n_c(\Phipg)=n$ for some integer $n$ (see Eqs.~(\ref{eq:ncGAD}) and~(\ref{eq:pEBn})).
 Notice that for all $\gamma$'s we have $p_{n}(\gamma)=p_{n}(1-\gamma)$ in agreement with Eq.~(\ref{ncnewid}). Furthermore, amplitude-damping channels, corresponding to $\gamma=0,1$ do not become EB after any number of iterations. In the inset figure, is shown the dependence of $\mu_c(\Phipg)$ and $\mu_c(\Phipg^2)$ on $p$, for $\gamma=1/3$. This functional vanishes as soon as the map becomes entanglement-breaking. As expected, for all $p$'s we find $\mu_c(\Phipg^2) \leq \mu_c(\Phipg)$ (see Eq.~(\ref{eq:mucn})).
}\label{fig:PEBn} 
\end{center}
\end{figure*}

By requiring the positivity of the determinant of the partially transposed Choi-Jamiolkowski state associated to this map \cite{augusiak}, 
we can determine the region of the parameter space $\{p,\gamma\}$ such that $\Psi_{p,\gamma}^n \in \EB$, see e.g. Fig.~\ref{fig:PEBn}.  
A direct calculation yields that $n_c(\Phipg)=n\geq 1$ if and only if 
\begin{eqnarray}\label{eq:ncGAD}
 p_{n} (\gamma) \leq p \leq p_{n-1} (\gamma)\;, 
 \end{eqnarray}
 where $p_0=1$ and 
 \begin{eqnarray}\label{eq:pEBn}
  p_{n} (\gamma)=1- \left(1-\frac{2}{1+\sqrt{1+4 \gamma (1-\gamma)}}\right)^{\frac{1}{n}}\;,
\end{eqnarray}
(see also~\cite{filippov} for $n=1,2$).

\begin{figure}[t]
\begin{center}
\includegraphics[height=0.7\columnwidth]{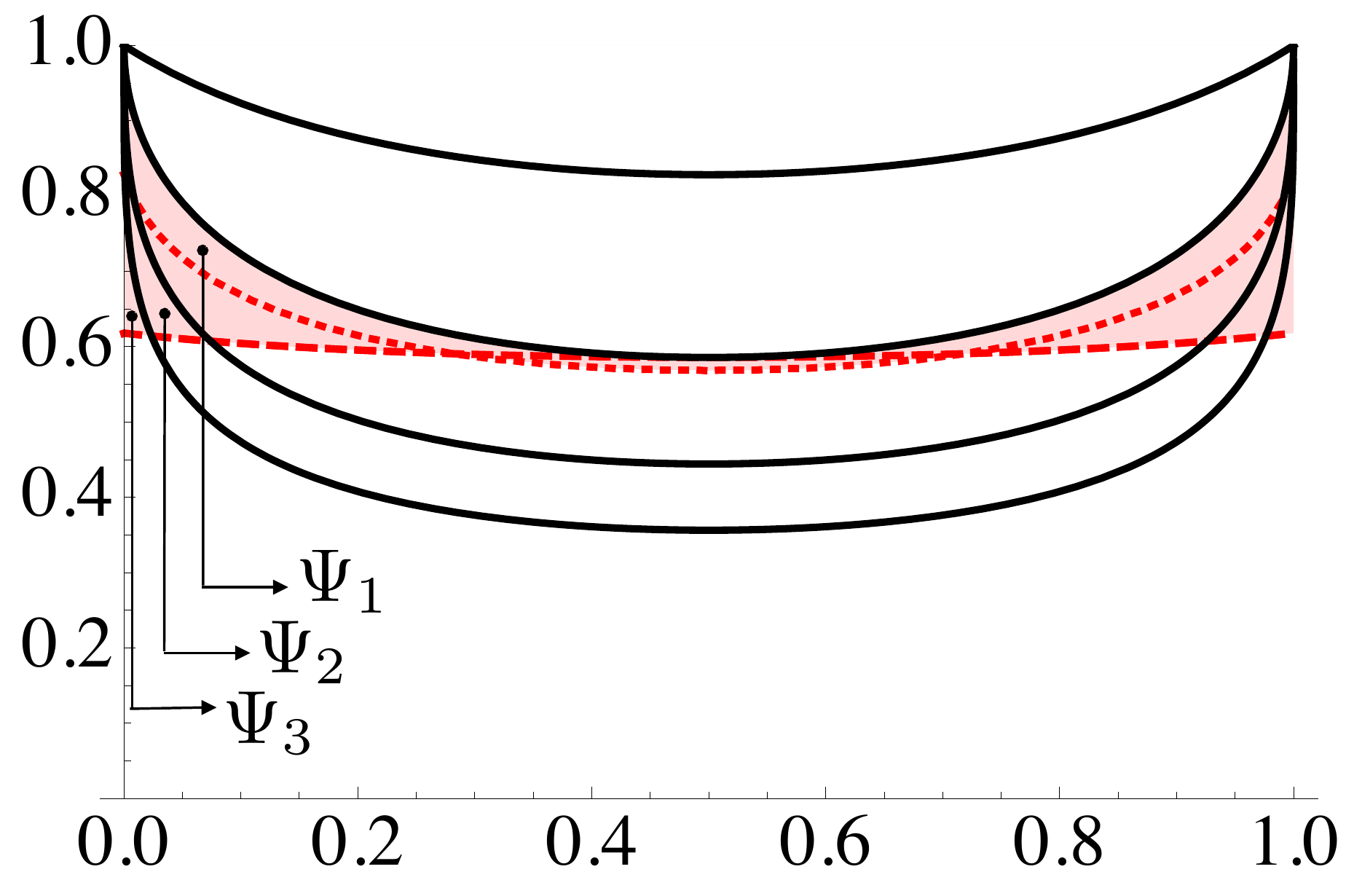}
\caption{(Color online) Amendable channels: the shaded region (red in online version) contains all points where (\ref{phiu}) holds for some ${\cal U}$.
The points $\Psi_1$, $\Psi_2$ and $\Psi_3$ are examples of CPT maps which are amendable of order $m=3$, $m=4$ and $m=5$ respectively (see text).}\label{fig:PhiRect}
\end{center}
\end{figure}

Starting from this observation we can now  construct  examples of amendable  channels, by looking for maps of the form
\begin{eqnarray}
\Phipg^{\mathcal{U}}=\Phipg \circ \mathcal{U} \label{phiu}\;,
\end{eqnarray}
with ${\cal U}$ being a unitary, such that 
$\Phipg^{\mathcal{U}}$ is  entanglement-breaking of oder $2$ with $n_c(\Phipg^{\mathcal{U}})=2$, 
while $\Phipg$ is not. Accordingly, some simple algebra allows us to write 
\begin{eqnarray} \label{ncphi22}
2= n_c(\Phipg^{\mathcal{U}}) < n_c(\Phipg) = n_c({\cal U}^\dag \circ \Phipg^{\mathcal{U}})\;,
\end{eqnarray} 
which provides an instance of Eq.~(\ref{ncphi}) with the filtering transformation ${\cal V}$ being now ${\cal U}^\dag$
(indeed for such examples we will have  $\Phipg^{\mathcal{U}} \circ  \mathcal{U}^\dagger\circ\Phipg^{\mathcal{U}} \notin \EB$ proving that we have amended  the action of the mapping $(\Phipg^{\mathcal{U}})^2$).  
The region of the parameter space $\{p, \gamma\}$  for which the inequality~(\ref{ncphi22}) holds for some ${\cal U}$ is shown in gray (red in online version) in Fig.~\ref{fig:PhiRect}.
A numerical investigation shows that it is given by the union of two areas. The first one  contains all points for which (\ref{ncphi}) holds with ${\cal U}$ being the Pauli rotation~(\ref{superss})  ${\cal S}_1$: this region
can be determined analytically and it is delimited from below by the condition 
 \begin{eqnarray}  \label{regione1} 
 p&\geq& \frac{1}{4
   (1-\gamma ) \gamma }
\Big[-\sqrt{4 (1-\gamma ) \gamma +1}\\&&+\sqrt{\left(1-2 \sqrt{1-4 (\gamma -1) \gamma }\right) (1-2 \gamma )^2+1}+1\Big]\;,  \nonumber 
 \end{eqnarray}
(dashed curve in the figure).
 The second one instead  contains all points for which  (\ref{ncphi})  holds for ${\cal U}= \mathcal{R}_2(\pi/2) \circ \mathcal{R}_1(\pi/2)$ with $\mathcal{R}_{j}$ is the super-operator associated with the rotations $\exp[-i (\pi/4) \sigma_{j}]$ (this has been characterized only numerically and it is delimited from below by the dotted line of the figure). 

Notice that for any given $m$ integer  the shaded area of Fig.~\ref{fig:PhiRect} has a not-null overlap with the region (\ref{eq:ncGAD})  associated with channels $\Phi_{p,\gamma}$ having $n_c(\Phi_{p,\gamma})=m$. 
A formal proof of this fact can be obtained by noticing that the boundary  determined by  Eq.~(\ref{regione1}) intercepts the vertical axis for   $\{\gamma,p\}=\{0,(\sqrt{5}-1)/2\}$ which,
according to Eq.~(\ref{eq:pEBn}), does not belong  the $m$-th region of Eq.~(\ref{eq:ncGAD}).
For these points  the rhs of Eq.~(\ref{ncphi22}) is explicitly equal to $m$, providing hence direct evidence of the existence of amendable channels of order $m$ 
(see Sec.~\ref{amendableorder}).

\begin{figure}
 \centering
\subfigure[] 
{\label{}\includegraphics[width=0.95\columnwidth]{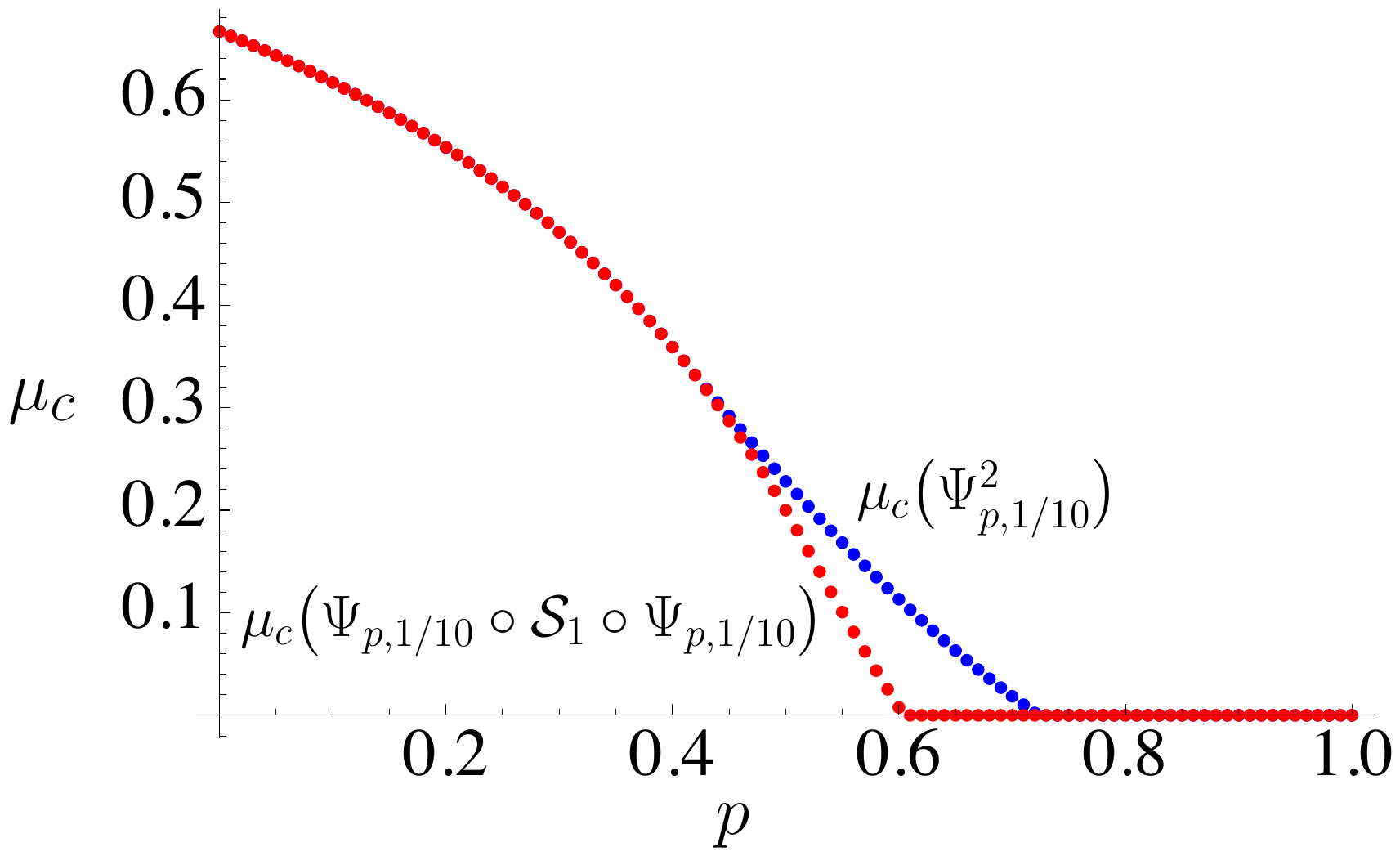}
}
 \hspace{5mm}
 \subfigure[] 
   {\label{}\includegraphics[width=0.95\columnwidth]{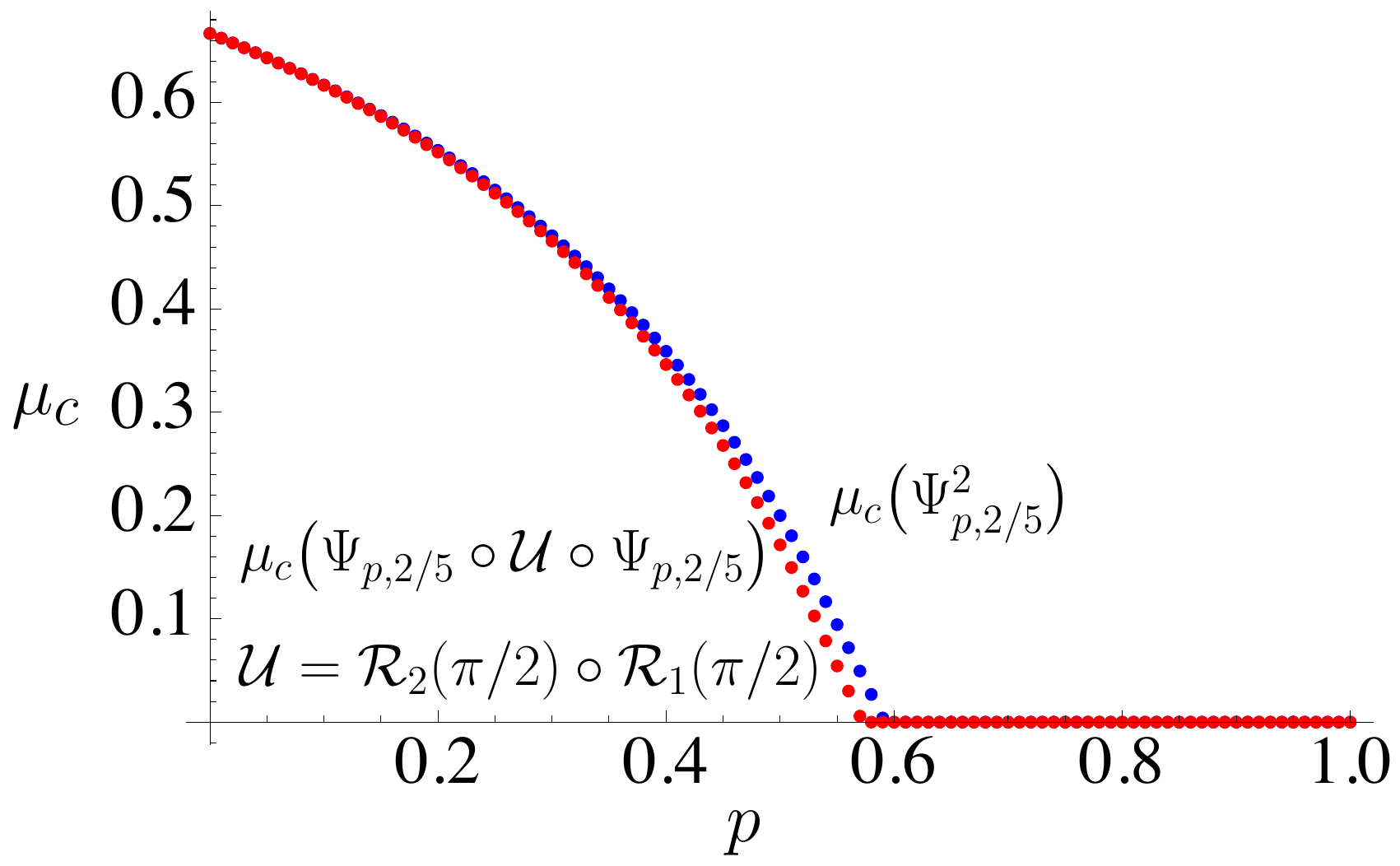}}
\caption{(Color online) Plot of  $\mu_c(\Phipg^2)$ and $\mu_c(\Phipg \circ \mathcal{U} \circ \Phipg)$, with $\mathcal{U}=\mathcal{S}_1$ for $\gamma=1/10$ (a) and $\mathcal{U}=\mathcal{R}_2(\pi/2)\circ\mathcal{R}_1(\pi/2)$ for $\gamma=2/5$ (b). For this particular choice of unitary maps we have that $\mu_c(\Phipg \circ \mathcal{U} \circ \Phipg)\leq \mu_c(\Phipg^2)$.}
\label{fig:PhiRPhi}
\end{figure}

In Fig.~\ref{fig:PhiRPhi} we finally compare the values of $\mu_c(\Phipg^2)$ with that of $\mu_c(\Phipg \circ \mathcal{U} \circ \Phipg)$ as functions of $p$ having fixed
${\cal U}={\cal S}_1$ and $\gamma =1/10$ (part (a)),  and $\mathcal{U}=\mathcal{R}_{2}(\pi/2)\circ\mathcal{R}_{1}(\pi/2)$ and $\gamma=2/5$  (part (b)). In both cases the
amended channels $\Phipg$ present an higher value of the functional $\mu_c$. 

\section{Gaussian Channels}\label{sec:gaussian}

In what follows we will show that behaviors similar to those seen in the previous section for finite dimensional systems, still hold also
for continuous variables. In particular we will discuss how $n_c$ can be used to
classify the set of Gaussian maps~\cite{gaussian} in a way which mimics what obtained for qubit channels.

Gaussian channels provide a mathematical description for the most common source of noise encountered in optical implementations, among which we have attenuation, amplification and thermalization processes \cite{gaussian}. They are CPT maps such that when operating on a Gaussian input state preserve its structure.
In particular, a state $\rho \in \mathfrak{S}({\cal H}_S)$ of a bosonic system with $f$ degrees of freedom, is Gaussian if its characteristic function $\phi_{\rho}(z)=\mathrm{Tr} [\rho W(z)]$ has a Gaussian form, where $W(z)$ is the  unitary Weyl operator defined on the real vector space $\mathbb{R}^{2f}$, $W(z):=\exp[i \vec{R}\cdot \vec{z}]$, with $\vec{R}=\{ Q_1, P_1, \ldots ,Q_f, P_f\}$ and $Q_i$, $P_i$ the canonical observables for the bosonic system. Gaussian maps $\Phi$ can be conveniently described in terms of the action of their duals $\Phi^*$ on the Weyl operator:
\begin{equation}
\Phi^*[W(z)]=W(Kz)\exp (i l^\top z-\frac{1}{2}z^\top \beta z)
\end{equation}
where $l \in \mathbb{R}^{2f}$, and $K$ and $\beta$ are $2f \times 2f$ matrices such that
\begin{equation}\label{eq:CPT}
\beta \geq \pm i [\Delta - K ^\top \Delta K]/2,\quad \end{equation}
$\Delta$ being the canonical symplectic form on $\mathbb{R}^{2f}$.
The above inequality is the necessary and sufficient condition which garantees the complete positivity of $\Phi$. A Gaussian channel is therefore characterized by the triplet $(K,l,\beta)$. The composition of two Gaussian maps, $\Phi=\Phi_2 \circ \Phi_1$, described by $(K_1,l_1,\beta_1)$ and $(K_2,l_2,\beta_2)$ respectively, is still a Gaussian map whose parameters are given by
\begin{eqnarray}\label{eq:composition}
\Phi_{2}\circ\Phi_1 \longrightarrow \left\{\begin{array}{l}K=K_1 K_2 \\
l=K_2^\top l_1 + l_2 \\
\beta=K_2^\top \beta_1 K_2+\beta_2.\end{array}\right.
\end{eqnarray}
Finally, a Gaussian map $\Phi$ is entanglement-breaking~\cite{HOLEVOEBG}  if and only if 
\begin{equation}\label{eq:EBTgauss}
\beta=\alpha+\nu,
\end{equation}
where
\begin{equation}\label{eq:EBTgauss2}
\alpha \geq  \frac{i}{2}\Delta, \quad \mbox{and} \quad \nu\geq \frac{i}{2}K^\top\Delta K.
\end{equation}
\subsection{One-mode channels}
For the case of continuous variables systems, one-mode Gaussian mappings ($f=1$) can be seen as the analogous of the qubit channels for finite dimensional systems. In this section we will quantify the noise level of two representative examples of this class of maps, i.~e. the attenuation and the amplification channels.
 By exploiting the conservation of the Gaussian character of these transformations under concatenation, and the entanglement-breaking conditions (\ref{eq:EBTgauss})-(\ref{eq:EBTgauss2}), in what follows we will focus on the classification of attenuation and amplification channels according to $n_c$, postposing the analysis of the addition via mixing to future works. 

\subsubsection{Attenuation channels}

One-mode attenuation channels $\Phi_{At}$ are Gaussian mappings such that:
\begin{equation}
\Phi_{At} \longrightarrow \left\{\begin{array}{l}K=k \openone\\l=0 \\ \beta=\left(N_0+\frac{1-k^2}{2}\right) \openone\end{array}\right.
\end{equation}
where $\openone=\left(\begin{array}{cc} 1 &0 \\ 0 & 1\end{array}\right)$, $k\in(0,1)$ and $N_0\geq0$, the latter condition deriving from the CPT  requirement~(\ref{eq:CPT}) on $\Phi_{At}$. 
From Eq.~(\ref{eq:EBTgauss}) it follows that $\Phi_{At} \in$  EB and $n_c(\Phi_{At})=1$ iff
\begin{equation}\label{EB1att}
\beta \geq \frac{i}{2}(1+k^2)\Delta \iff N_0 \geq k^2.
\end{equation} 
In order to generalize the condition above to $n_c({\Phi_{At}})=n$ with $n\geq2$, let us start by studying the composite map ${\Phi_{At}}^2 :=\Phi_{At} \circ \Phi_{At} $. It is given by
\begin{equation}
{\Phi_{At}}^2 \longrightarrow \left\{\begin{array}{l}  K_2=k^2 \openone\\l_2=0 \\ \beta_2=(1+k^2)\left(N_0+\frac{1-k^2}{2}\right) \openone\,.\end{array}\right.
\end{equation}
Notice that if we define
\begin{equation}
k_2:=k^2 \leq 1     \quad \mbox{and} \quad(N_0)_2:=N_0 (1+k^2) \geq 0
\end{equation}
we get 
\begin{equation}
   \beta_2=\left((N_0)_2 + \frac{1-k_2^2}{2} \right) \openone\,,  
\end{equation}
which plays the same role of $\beta$ for $\Phi_{At}$.
It immediately follows that ${\Phi_{At}}^2 \in$ EB if and only if
\begin{equation}
\beta_2 \geq  \frac{i}{2} (k_2^2+1)\Delta \implies (N_0)_2 \geq k_2^2\end{equation}
(see Eq.~(\ref{EB1att})), that is $n_c(\Phi_{At})=2$ for 
\begin{equation}
 \frac{k^4}{1+k^2} \leq N_0 \leq k^2.
\end{equation}
We can now generalize these results to the case $n_c(\Phi_{At})=n\geq2$. From Eq.~(\ref{eq:composition}) we get
\begin{equation}
{\Phi_{At}}^n\longrightarrow \left\{\begin{array}{l}K_n=k_n \openone\\l_n=0 \\ \beta_n=\left((N_0)_n+\frac{1-k_n^{2}}{2}\right) \openone\,,\end{array}\right.
\end{equation}
where
\begin{equation}\label{eq:knNon}
k_n:=k^n \leq 1 \quad \mbox{and}  \quad (N_0)_n:=N_0\sum_{j=0}^{n-1}k^{2j} \geq 0\,.
\end{equation}
It follows that ${\Phi_{At}}^n \in$ EB iff $(N_0)_n \geq k_n^2$, and $n_c(\Phi_{At})=n$ if and only if 
\begin{equation}\label{eq:EBnAtt}
 \frac{k^{2n}}{\sum_{j=0}^{n-1}k^{2j}} \leq N_0 \leq \frac{k^{2(n-1)}}{\sum_{j=0}^{n-2}k^{2j}}.
\end{equation}

\subsubsection{Amplification channels}
\begin{figure}[t]
\centering
\includegraphics[width=1\columnwidth]{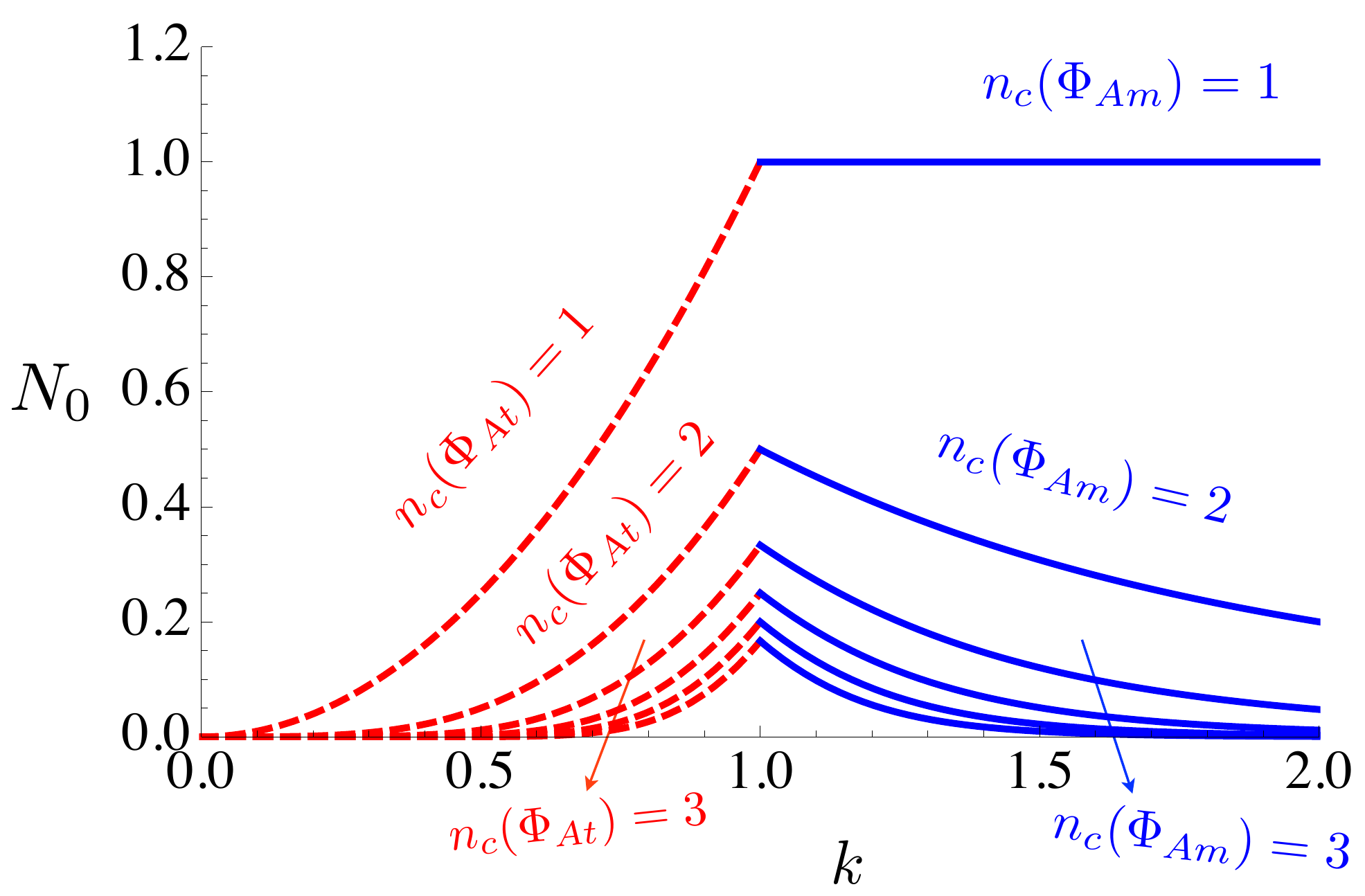}
\caption{(Color online) Boundaries of the regions in the parameter space  $\{k,N_0\}$ such that $n_c(\Phi_{At})=n$ (dashed red lines), and $n_c(\Phi_{Am})=n$ (solid blue lines) for $n=1, \ldots, 6$.}
\label{fig:attenuationANDamplification}
\end{figure}

One-mode amplification maps are very similar to the attenuation channels, where now $k\geq1$ and
\begin{equation}
\Phi_{Am} \longrightarrow \left\{\begin{array}{l} K=k \openone\\ l=0 \\ \beta=\left(N_0+\frac{k^2-1}{2}\right) \openone. \end{array}\right.\end{equation}
From Eq.~(\ref{eq:EBTgauss}) we have that the entanglement-breaking condition for this class of maps is 
\begin{equation}\label{eq:EBampl}
\beta \geq \frac{i}{2}(1+k^2)\Delta \iff N_0 \geq 1.
\end{equation}
For $n\geq2$ one has
\begin{equation}
{\Phi_{Am}}^n\longrightarrow \left\{\begin{array}{l}K_n=k_n \openone 
\\l_n=0
\\ \beta_n=\left((N_0)_n+\frac{k_n^{2}-1}{2}\right) \openone,
\end{array}\right.
\end{equation}
with $k_n$ and $(N_0)_n$ given by Eq.~(\ref{eq:knNon}). It follows that the composite map
${\Phi_{Am}}^n$ is entaglement-breaking iff $(N_0)_n \geq 1$, and $n_c(\Phi_{Am})=n\geq2$ for
\begin{equation}\label{eq:EBamplN}
\left(\sum_{j=0}^{n-1}k^{2j}\right)^{-1} \leq N_0 \leq \left(\sum_{j=0}^{n-2}k^{2j}\right)^{-1}.
\end{equation}
Analogously to what done in Fig.~\ref{fig:PEBn} for the case of generalized amplitude-damping channels, in Fig.~\ref{fig:attenuationANDamplification} we propose a scan of the parameter space $\{k,N_0\}$ for attenuation and amplification maps according the $n_c$ criterion by plotting the boundaries of the regions such that $n_c(\Phi_{At})=n$ (dashed red lines), and $n_c(\Phi_{Am})=n$ (solid blue lines) for $n=1, \ldots, 6$. Notice that, the region such that $\Phi_{At/Am}\in \EB^n$ grows with $n$, as it is limited from below by the curves
\begin{eqnarray}
\left\{\begin{array}{l} N_0=k^2,\;\;\;\;\quad\quad n=1\\N_0=\frac{k^{2n}}{\sum_{j=0}^{n-1}k^{2j}},\; n\geq2 \\  \end{array}\right.\end{eqnarray} and \quad \begin{eqnarray}\left\{\begin{array}{l}N_0=1, \,\qquad\qquad\qquad  n=1\\ N_0=\left(\sum_{j=0}^{n-1}k^{2j}\right)^{-1},n\geq2 \end{array}\right. \end{eqnarray}
for attenuation and amplification channels, respectively. See Eqns.~(\ref{EB1att}), (\ref{eq:EBnAtt}), (\ref{eq:EBampl}) and (\ref{eq:EBamplN}).

\section{Conclusions}\label{par:Summary and Conclusions}

In our analysis we have introduced two new functionals ($\mu_c$ and $n_c$) that can be used to quantify the noise level of a map. These have been characterized in terms of general properties
and have been explicitly evaluated for some class of channels. 
Along the way we have also introduced the notion of amendable channels showing how there exist maps which can be prevented from becoming entanglement-breaking after subsequent applications 
by interposing some extra transformations.  For the sake of simplicity in our study we have only considered the case in which the \emph{same} unitary transformation is interposed between successive channel
uses. This however is not the only possibility and it is an interesting question of quantum control to determine what are the optimal operations one has to perform in order to guarantee that after certain number
of reiterations entanglement will not be destroyed in the system.

We finally notice that when introducing the functionals $n_c$ and $\mu_c$ we started from the identification of the  entanglement-breaking channels as a benchmark set to evaluate the noise level of other quantum transformations 
(the choice
being operationally motivated by the extreme deteriorating effects that  EB maps have on the system).
Again this however is not the only possibility. For instance another reasonable choice is to  replace
the entanglement-breaking channels with the set  formed by  the Positive Partial Transpose (or {\em binding}) maps~\cite{PPT} producing  new noise measures $\mu_c^{(PPT)}$ and $n_c^{(PPT)}$ for $\Phi$.
We remind that a channel is said to be PPT 
 if when extended on an ancilla A the only entanglement one can find in the output of $S+A$ is  non-distillable~\cite{dist}. Similarly to the case of EB maps, 
 also the PPT set is stable under convex convolution and
 iteration: therefore $\mu_c^{(PPT)}$ and $n_c^{(PPT)}$ will again assume the minimum allowed values on the benchmark set guarantying that they
 are well defined quantities for all $\Phi$.  Now, since for qubits the entanglement is always
 distillable~\cite{h4}, when $S$ has dimension 2 we have $\mu_c^{(PPT)}(\Phi)=\mu_c(\Phi)$ and $n_c^{(PPT)}(\Phi)=n_c(\Phi)$ (a property we explicitly exploited when computing the values of $\mu_c$ and $n_c$ for qubit channels). On the contrary when $S$ has larger dimensions, 
 $\mu_c^{(PPT)}(\Phi)$ and $n_c^{(PPT)}(\Phi)$ need not to reduce to 
 $\mu_c(\Phi)$ and  $n_c(\Phi)$ yielding a qualitatively new way of gauging the noise level of $\Phi$.

\acknowledgments 
This work was supported by MIUR through FIRB- 
IDEAS Project No. RBID08B3FM.
 
\appendix
\section{A convexity property of $\mu(\Phi;{{\rho_0}})$ }\label{appendixA}
In this appendix we show a convexity rule for the functional $\mu(\Phi; {{\rho_0}})$ with respect to its second argument.

Consider a generic statistical ensemble  $\{ p_j; \Phi^{\rho_j}_\DEP\}$ of   completely depolarizing maps $\Phi^{\rho_i}_\DEP$ distributed according to the probabilities $p_j$. We will prove that $\mu(\Phi; {{\rho_0}})$ is convex with respect to its second argument:
 \begin{equation}\label{eq:convmu1}
 \mu(\Phi; \sum_{i}p_i {\rho_i}) \leq \tilde{\mu} \leq  \sum_{i}p_i \, \mu_{\DEP}^{j}\,,
 \end{equation} 
 where $\mu_{\DEP}^{j}:= \mu\left(\Phi; {{\rho_j}}\right)$, $\rho_j \in \mathfrak{S}({\cal H}_S)$ and
 \begin{equation}
 \tilde{\mu}=\left(\sum_{i} \frac{p_i}{\mu_{\DEP}^{j}}\right)^{-1}\,.
 \end{equation}
In order to prove the first inequality we define the probabilities $\tilde{q}_i:=\frac{p_i}{\mu_{\DEP}^{i}}/(\sum_{j} \frac{p_j}{\mu_{\DEP}^{j}})$ and notice that 
\begin{equation}
\sum_j \tilde{q}_j  \Gamma^{\Phi}_{\rho_j,\mu_{\DEP}^{j}}= (1-\tilde{\mu}) (\Phi \otimes I)[ \psi_{+}]+\tilde{\mu}
 \left(\sum_{i}p_i \rho_i \otimes {\openone}/{d} \right )\nonumber  
\end{equation} 
is separable. The second inequality in~(\ref{eq:convmu1}) is equivalent to
\begin{equation}
\prod_\ell{\mu_{\DEP}^{\ell}}\leq \left(\sum_{i}p_i \prod_{j\neq i} {\mu_{\DEP}^{j}} \right)\left(\sum_{k}p_k \, \mu_{\DEP}^{k}\right)\end{equation}
and can be easily proved applying the normalization condition $\sum p_i=1$. 

\section{A  decomposition for unital qubit maps}\label{sec:KrausUnitalQubit}

In this appendix we verify the commutation rule~(\ref{commuting}) by noticing that 
 for unital qubit maps in canonical form, $\Phil$  there exists a decomposition in terms of the
 super-operators~(\ref{superss}), i.e. 
\begin{equation} \label{phirho}
\Phil=\sum_{0\leq i \leq 3}p_i \mathcal{S}_i\;,
\end{equation}
with $p_i$ being real quantities (as a matter of fact $p_i$ can be shown to be probabilities~\cite{ruskai}: here however we will not need this property). From this then~(\ref{commuting}) simply follows from the fact 
that  for all $i,j$ we  have $\mathcal{S}_j \circ \mathcal{S}_i \circ  \mathcal{S}_j  = \mathcal{S}_i$.

To see Eq.~(\ref{phirho}) we expand  the input state  $\rho$   in the Bloch sphere formalism as $\rho=(1+\vec{v}\cdot \vec{\sigma})/2$ and use the fact that 
\begin{equation}\label{b2}
\Phil[\rho]=\frac{1}{2}\sum_{0\leq i \leq 3} \lambda_i v_i \mathcal{\sigma}_i\,, 
\end{equation}
where we set $\lambda_0 =v_0=1$.
On the other hand,  if we apply the super-operator $\sum_{0\leq i \leq 3}p_i \mathcal{S}_i$ to $\rho$ we get
\begin{equation}\label{b3}
\sum_{0\leq i \leq 3}p_i \mathcal{S}_i[\rho]=\frac{1}{2}\sum_{0\leq i,j \leq 3} p_i  M_{i j} v_j \; \sigma_j\;,
\end{equation}
where  $M_{ij}$ are elements of  the $4\times 4$ invertible symmetric real matrix 
\begin{eqnarray}
M : = \left[\begin{array}{rrrr}
1 & 1&1&1 \\
 1 & 1&-1&-1 \\
 1 &- 1&1&-1 \\
 1 & -1&-1&1 \\
\end{array} \right]\;,
\end{eqnarray}
defined by the relation $M_{ij}\sigma_j={\cal S}_i(\sigma_j) = \sigma_i \sigma_j \sigma_i$. By equating Eqs.~(\ref{b2}) and (\ref{b3}) we conclude that Eq.~(\ref{phirho}) holds if and only if
 \begin{equation}
 \vec{p}=M^{-1} \vec{\lambda}\;,
\end{equation}
where here $\vec{p}$ and $\vec{\lambda}$ stand for the 4-dimensional vectors $(p_0, \ldots,p_3)$ and 
$(\lambda_0, \ldots, \lambda_3)$, respectively.


\begin{thebibliography}{99}
\bibitem{NOTA}
In defining the Choi-Jamiolkowski state~(\ref{eq:CJ}) for the qubit case we identify 
$|\psi_+\rangle$ with the Bell triplet state $(|00\rangle+ |11\rangle)/\sqrt{2}$ whose associated density
matrix is indeed invariant under $\mathcal{S}_j\otimes {\mathcal S}_j$.
\bibitem{NOTA3} This is a consequence of the fact that the matrix $D$ associated to $\Phi_\lambda$ via~(\ref{correspondence}) is diagonal. Hence 
$|D^n|=|D|^n$ so that $\| D^n\|_1=\mbox{Tr}[|D|^n] = \mbox{Tr}[ \Lambda^n]$ where in the last identity we used the fact that $|D|$ has the same eigenvalues
of $\Lambda$. From Eq.~(\ref{COMPUTENC}) it then follows that $n_c(\Phi_p)= n_c(\Phi_\lambda)$.
\bibitem{NOTA2} Of course for $n=1$ one has that $\Phi\in \EB \Longleftrightarrow \Phi_p \in \EB$ which implies $n_c(\Phi)=1 \Longleftrightarrow  n_c(\Phi_p)$.
\end{thebibliography}
\end{document}